\documentclass[12pt,preprint]{aastex}

\usepackage{amssymb}

\shorttitle{Optically Invisible Radio And X-Ray Sources}
\shortauthors{Higdon et al.}

\begin{document}

\title{ Spitzer Observations of Optically ``Invisible''
        Radio And X-Ray Sources: 
        High Redshift AGN}

            
\author{ J. L. Higdon\altaffilmark{1}, 
         S. J. U. Higdon\altaffilmark{1},
         D. W. Weedman\altaffilmark{1},
         J. R. Houck\altaffilmark{1},  
         E. Le Floc'h\altaffilmark{2},
         M. J. I. Brown\altaffilmark{3}, 
         A. Dey\altaffilmark{4},  
         B. T. Jannuzi\altaffilmark{4},
         B. T. Soifer\altaffilmark{5},
      \& M. J. Rieke\altaffilmark{2}}

\altaffiltext{1}{Astronomy Department, Cornell University, Ithaca, NY 14853}
\altaffiltext{2}{Steward Observatory, University of Arizona, Tucson, AZ 85721}
\altaffiltext{3}{Astronomy Department, Princeton University, Princeton, NJ 85726 }
\altaffiltext{4}{National Optical Astronomy Observatory, Tucson, AZ 85726 }

\begin{abstract}
We have combined a survey at 24 $\mu$m to 0.3 mJy with
the Multiband Imaging Photometer (MIPS) on the Spitzer Space
Telescope, a 20 cm A-array Very Large Array survey covering
0.5 deg$^{2}$, and an existing 172 ks Chandra X-Ray Observatory 
exposure to investigate the nature of optically faint radio and 
X-ray sources in the NOAO Deep Wide-Field Survey (NDWFS) in Bo\"otes.  
We find little overlap between the radio and infrared
selected populations: only 9\% of the infrared sources are detected in
the radio and only 33\% of the radio sources are detected in the
infrared.  Thirty-six (10\%) of the 377 compact radio sources lack 
optical counterparts in the NDWFS $B_W$, $R$, \& $I$ images.  We refer
to these objects as optically invisible radio sources (OIRS).
Only four (13$\%$) of the thirty-one OIRSs observed with MIPS are detected 
at 24 $\mu$m. Comparisons of the radio and infrared properties of 
the OIRSs with various galaxy spectral energy distributions demonstrate 
that most of these sources are powered by AGN rather than starbursts.
Similarly, eleven X-ray sources observed by both MIPS and the VLA are 
classified as optically invisible X-ray sources (OIXS). None are detected 
at 24 $\mu$m or 20 cm. All seven OIXSs detected in Chandra's 0.5-2 keV band have
infrared to X-ray flux ratios consistent with their luminosity being
dominated by an unobscured AGN. From these results we conclude 
that both the optically invisible radio and X-ray source populations 
are primarily AGN, relatively unaffected by dust and most likely at z $>$ 1.  
No OIRSs are detected in X-ray emission and no OIXSs are detected at 20 cm.  
However, given the wide range in radio and X-ray properties of known AGN
and the size of our samples, this lack of overlap does not necessarily 
imply different AGN source populations.
\end{abstract}

\keywords{ galaxies: high-redshift -- galaxies: starburst --
           galaxies: AGN -- infrared: galaxies -- 
           radio: galaxies -- X-ray: galaxies }

\section{Introduction}

Determining the star formation density of the Universe beyond z$\sim$1
and the contribution made by AGN to the luminosity evolution of
galaxies are two fundamental goals of observational cosmology.
Addressing these questions requires observations at wavelengths
capable of locating intrinsically luminous sources that are either 
heavily obscured
by dust, or so distant that their rest-frame optical wavebands
fall below the cutoff for intergalactic hydrogen absorption (i.e.,
the Lyman limit). Locating and understanding such obscured sources is a
fundamental objective of the {\em Spitzer Space Telescope} 
(Spitzer; Werner et al. 2004)
mission\footnotemark[1], \footnotetext[1]{The Spitzer Space Telescope
is operated by JPL, California Institute of Technology for the
National Aeronautics and Space Administration.  Information on Spitzer
can be found at http:$//$ssc.spitzer.caltech.edu$/$} which is capable
of identifying z $>$ 1 galaxy populations that are luminous at
infrared wavelengths but faint optically due to the effects of dust.  
Starbursts and AGN are expected to be
powerful emitters at radio wavelengths due to large populations of
radio-luminous supernova remnants for starbursts and accretion disk
phenomena for AGN. Indeed, sensitive radio surveys routinely
find populations of sub-mJy point sources with either very faint
optical counterparts or no counterparts at all (e.g., Richards et
al. 1999; Fomalont et al. 2002).  Whether these
sources represent dust enshrouded starburst galaxies at z $\sim$ 1-3
or radio loud AGN at z $>$ 4 has yet to be determined conclusively.  
This suggests as a strategy for locating heavily
obscured starbursts and AGN at high redshifts the definition of
infrared and radio source populations that are either extremely
faint at all optical bands or ``invisible'', i.e., objects that 
can not be distinguished from the background in deep optical 
images.

The NOAO Deep Wide-Field Survey\footnotemark[2] (NDWFS; Jannuzi \& Dey 1999)
\footnotetext[2]{The NOAO Deep Wide-Field Survey is supported
by the National Optical Astronomy Observatory, which is 
operated by AURA, Inc., under a cooperative agreement with the
National Science Foundation. Information on the NDWFS can 
be found at http:$//$www.noao.edu$/$noao$/$noaodeep$/$.}
provides multiband optical imagery to very faint magnitudes over
large survey areas, and so is uniquely suited for comparison to other
surveys in search of sources that are optically invisible. In
particular, the NDWFS field in Bo\"otes covers $\sim$9.3 deg$^{2}$ to
sensitivities in $B_W$, $R$, \& $I$ of approximately
26.5, 25.5, \& 24.7 (Vega) magnitudes respectively.\footnotemark[3]
\footnotetext[3]{These represent the 50$\%$ completeness limits in 
the source catalogs for the area involved in this study. Similar
sensitivities are expected throughout the survey.}  To explore the
mid-infrared characteristics of this data set, we have surveyed most
of the Bo\"otes field at wavelengths of 24, 70, \& 160 $\mu$m with
Spitzer's Multiband Imaging Photometer (MIPS; Rieke et al. 2004). 
At 24 $\mu$m the point source sensitivity is 0.3 mJy (5 $\sigma$) 
and the angular resolution is 5.5\arcsec (FWHM). We will present 
other results of this infrared survey in more detail in future 
publications, although the survey and analysis techniques have been 
described by Papovich et al. (2004).

Deep and high resolution radio data are important for a full
understanding of the nature of the sources detected by MIPS. 
This follows from the well established tight correlation between
infrared and radio flux densities in star forming galaxies (de Jong
et al. 1985; Helou et al. 1985), which is a consequence
of the life (re-radiation of UV photons by dust) and death (SNe remnants) 
of massive stars.  As a result, the ratio of infrared to radio flux 
densities provides a means of distinguishing
starbursts from AGN.  Moreover, the sub-arcsecond 
positional accuracy attainable with radio interferometric surveys allows
confident matching of sources discovered in infrared surveys,
which can have poorer spatial resolution. This is especially important
if a source is indeed optically invisible. 
For these reasons, we conducted sensitive and high angular resolution 
observations of a 0.5 deg$^{2}$ subregion within the Bo\"otes field
with the {\em Very Large Array} (VLA) at 20 cm.  These observations were
designed to detect dusty starburst galaxies with 24 $\mu$m flux
densities greater than 0.75 mJy based on their expected
spectral energy distributions and likely redshift range.\footnotemark[4]
\footnotetext[4]{For these estimates we used the starburst infrared/radio 
correlation of Yun et al. (2001) together with K-corrections from the SED
templates of Lagache et al. (2002).}
Although the MIPS 24 $\mu$m survey reaches a fainter limit than
this, we chose this infrared limit for matching radio observations
because we were primarily interested in selecting sources bright enough
to obtain spectra with the Infrared Spectrograph (IRS; Houck et al. 2004) 
on Spitzer (i.e., Higdon et al. 2004, Houck et al. 2005). 

Our survey region was chosen to overlap the
northern Large Area Lyman Alpha survey field (LALA; Rhoads et al. 2000),
which includes an extremely deep (172 ks) {\em Chandra
X-Ray Observatory} integration by Wang et al. (2004), 
centered at RA = 14$^{h}$ 25$^{m}$ 37.79$^{s}$ 
and Dec = +35$^{\circ}$ 36$\arcmin$ 00.2$\arcsec$ and covering
0.1 deg$^{2}$.  The X-ray point source sensitivities are  1.5
$\times$ 10$^{-16}$ erg s$^{-1}$ cm$^{-2}$ in the ``soft'' 0.5-2.0
keV band and 1.0 $\times$ 10$^{-15}$ erg s$^{-1}$ cm$^{-2}$ in
the ``hard'' 2.0-7.0 keV band (both 5~$\sigma$).  Only the North and
South Chandra Deep Fields (2 and 1 Ms integrations respectively) are
substantially deeper. As a result, there was minimal 
overlap with the much more extensive 20 cm survey of the Bo\"{o}tes field by 
de Vries et al. (2002) using the Westerbork Synthesis Radio Telescope 
(WSRT).  In this paper we will use the combined X-ray, optical, infrared, 
and radio catalogs to investigate the nature of {\em Optically Invisible
Radio and X-ray Sources} (OIRS \& OIXS) in our radio survey area. 
These are defined to be X-ray and compact radio sources that have 
no apparent counterparts in the NDWFS $B_W$, $R$, \& $I$ images. 
In addition, we wish to see how the 
properties of these two optically invisible populations compare with 
one another.

The outline of this paper is as follows: we first discuss the VLA 
observations and data reduction in $\S$2.  This is followed by a 
brief discussion of the radio source catalog in $\S$3 and our criteria 
for matching sources from the MIPS, Chandra, and VLA surveys. We then 
describe the infrared, radio, and X-ray properties of the optically invisible
radio and X-ray sources ($\S$4) in order to constrain the dominant
source of their luminosity.  Finally, before summarizing in $\S$6,
we compare the average space densities of the OIRSs and 
OIXSs with those of other galaxies, including quasars, and also
consider whether OIRSs and OIXSs are drawn from the same population in $\S$5.
Throughout this paper, we have assumed a flat $\Lambda$CDM cosmology
with $\Omega_{\rm M}$=0.27, $\Omega_{\rm \Lambda}$=0.73, and a
Hubble constant of 71 km s$^{-1}$ Mpc$^{-1}$.

\section{Radio Observations And Reduction}

	We observed three overlapping fields within the NDWFS in
Bo\"otes with the VLA in the A-configuration at 20 cm.  These
fields, labeled North, East, and South, are shown in Figure 1, which
also depicts the Chandra survey field of Wang et al. (2004).  The
field center coordinates and other details of the observational setup
are given in Table 1.  Spectral line mode was used to minimize
bandwidth smearing and facilitate the flagging of narrow band radio
interference.  The correlator was configured to provide two sets of
seven channels (3.125 MHz separation), each centered at 1.3649 and
1.4351 GHz.  Both left and right handed circular polarizations were
measured at the two frequencies.  Each field was observed in 40 minute
blocks, inter-spaced with 2 minute observations of the nearby secondary
calibrator 1416+347 (S$_{\rm 1.435~GHz}$ = 1.96 $\pm$ 0.01 Jy) to
correct drifts in the antenna gains and phases.  Flux and bandpass
calibration were accomplished with longer observations of the primary
flux calibrator 3C 48 (S$_{\rm 1.435~GHz}$ = 15.01 $\pm$ 0.01 Jy,
Baars et al. 1977) made at the start of each night.  On-source
integrations totaling 5.5, 4.9, and 5.4 hours were achieved at the
North, East, and South positions, respectively. 

	The visibilities were processed using routines in 
NRAO's Astronomical Imaging Processing System (AIPS)
software package.  A standard antenna based calibration was applied to
each data set after excising a very small fraction of the data 
that was corrupted by radio interference.
The routine IMAGR was used to map the primary beam with the correct
w-phase terms (DO3DIMAG=1).  Images were made using a robust weighting
parameter of 1.1, which improved the beam shape at the expense of an
$\sim$7$\%$ increase in noise.  Radio sources from the FIRST catalog
(Becker et al. 1995) that were located outside the primary beams yet
capable of contributing significant side lobes were also cleaned.
Self-calibration was helpful in improving the antenna based
calibration, particularly in the South field, which is dominated by a
double-lobed radio source (87GB 142337.6+351136). After convolving to
a common resolution of 1.40\arcsec ~FWHM, the three final images were placed
on a common grid and combined.  The resulting rms at the three
field centers is $\sim$15 $\mu$Jy beam$^{-1}$,  which is in good
agreement with the expected noise levels using the array parameters
listed in the VLA Observational Status Summary. Since we have imaged
only three array pointings, significant variations
in sensitivity exist across the survey field. Close to the field
centers where the sensitivity is the highest we can detect point
sources as faint as $\sim$80 $\mu$Jy.  However, the 
faintest point source we are capable of detecting throughout the
entire survey is $\sim$200 $\mu$Jy. For this reason we do not attempt
to produce a radio source catalog that is complete to a given flux density
limit. This does not affect our ability to calculate
flux density ratios and limits for the radio, infrared, and 
X-ray sources.  However it will impact our estimates of
surface densities.

\section{Source Identification And Matching} 

	We restricted our radio survey to where the sensitivity
in the combined pointings exceeded 40$\%$ of the pointing
centers, i.e., where the 1~$\sigma$ noise is less than
37 $\mu$Jy beam$^{-1}$. This corresponds to a total area of 0.5 
deg$^{2}$. We used SExtractor (Bertin \& Arnouts 1996) to identify 
radio sources in our survey rather than the AIPS routine SAD or its 
variants, even though SExtractor is more commonly used to find and catalog 
sources in optical images. This was because of the package's flexibility 
and our experience using it. After first smoothing the input radio image
with a Gaussian profile matching the synthesized beam, SExtractor 
identified groups of at least five contiguous pixels, each exceeding 
the locally defined 2.5~$\sigma$ background fluctuations.
From this first cut of source identification, only those with a 
peak pixel value exceeding five times the local background rms were 
classified as source candidates. These were checked visually, 
and the small fraction found to be detections of residual sidelobes 
or radio source structure (e.g., hot spots in jets) were deleted from the
catalog.  A total of 392 sources satisfied these selection criteria, 
of which 377 are point-like or marginally resolved.  These objects 
are classified as ``compact'' objects in the radio source catalog.
Flux densities were measured using
the AIPS task IMFIT with the final primary beam
corrected image. We determined fluxes for extended objects using the 
AIPS task BLSUM.

We estimated the number of false detections in our survey by running
SExtractor with the same settings on the radio image multiplied by
-1. Assuming that the background noise is symmetric with respect to
the mean, the number of sources detected in the reversed image
should equal the number of false positives.  Only three negative ``sources''
were found that were not associated with side-lobes, and all of these were
$<$ 5.5 times the local rms.
As a check of this analysis, we calculated the number of independent
beams in the radio survey and then estimated the number of $\geq$ 5 $\sigma$
detections one would expect from random fluctuations.  For 5 $\times$ 10$^{6}$
beams we expect only two spurious 5 $\sigma$ detections and none at the 6 $\sigma$
level.  We therefore expect minimal contamination of our radio catalog by
false sources, with two to three of the nearly four-hundred detections being
spurious.

We compared the coordinates of the compact sources with
the NDWFS optical catalog.\footnotemark[5] \footnotetext[5]{Catalogs
were derived using the third data release (DR3) version of the NDWFS 
optical images (Jannuzi et al. in preparation, 
http:$//$www.noao.edu$/$noao$/$noaodeep).} It was found that two-thirds of 
the radio sources had an optical counterpart within 0.3\arcsec. Because the
great majority of these identifications are real, this value was
taken as the 1 $\sigma$ relative positional accuracy.  Similarly,
the measured 1~$\sigma$ scatter in $\Delta$RA and $\Delta$Dec between 
the 24 $\mu$m sources and their nearest optical counterparts is 0.3\arcsec~
for F$_{\rm 24 \mu m}$ $\ge$ 0.75 mJy. The positional accuracy was
somewhat worse ($\sim$1$\arcsec$) for the MIPS sources near 0.3 mJy. 
The positional accuracy of the ``soft'' band X-ray sources listed in 
Table 1 of Wang et al. (2004) ranges from 0.3 to 3.5\arcsec, but is typically 
1\arcsec. We will adopt this as the X-ray source positional accuracy.

The identification of optically invisible radio sources was a two
step process.  First we correlated the positions of all 377 compact
radio sources with the NDWFS $B_W$, $R$, and $I$-band catalogs,
stipulating that OIRS candidates must not have a cataloged optical
counterpart centered within a 1.5\arcsec~radius (5 $\sigma$).  For the candidates
passing this test, we inspected the NDWFS images at these positions
to verify that no optical source was visible in any band.  A small number 
of candidates without apparent optical counterparts but close to the edges 
of field galaxies were also rejected. This process yielded thirty-six OIRSs, 
which we list in Table 2.  The identification of optically invisible
X-ray sources was carried out in a similar manner, starting with the
thirteen sources in Table 1 of Wang et al. (2004) with positional uncertainties 
less than 2\arcsec ~that lack $B_W$, $R$, \& $I$ counterparts.

To determine if the OIRSs and OIXSs had infrared counterparts, 
we then compared their positions with the MIPS 24 $\mu$m catalog
(F$_{\rm 24 \mu m}$ $\ge$ 0.3 mJy), calling a radio and infrared source
matched if their positions differed by less than 2.5\arcsec.  
Infrared fluxes or upper-limits for the OIRSs are given in Table 2.
Similarly, we adopted as criteria for matching radio to X-ray and X-ray to 
infrared sources that their coordinates agree to $<$ 2\arcsec.

\section{Results And Analysis}

\subsection{Properties Of The OIRS Population}

We identified a total of thirty-six OIRSs over the 0.5 deg$^{2}$ survey area,
a number representing 10$\%$ of the total compact radio source population.
Table 2 lists their radio catalog number, source name incorporating J2000 
coordinates, and physical properties.  Nearly all (86$\%$) are detected
above the 5.5 $\sigma$ level, where we expect no spurious sources. 
Figure 2 shows 12$\arcsec \times$12$\arcsec$ 
sub-images from the NDWFS $B_W$, $R$, and $I$ data centered on five of the OIRSs, 
along with the corresponding sub-images from the MIPS 24 $\mu$m survey 
field.  The radio continuum source structure for each is shown contoured 
on the I-band sub-images, with the lowest contour representing the local 
3 $\sigma$ noise level in the radio image.

The precise magnitude limit for an OIRS will depend on the
position of the source in the NDWFS field.  We therefore derived
$B_W$, $R$, and $I$ upper-limits for the 31 OIRS within 2$\arcsec$
diameter apertures centered on their positions using the appropriate
NDWFS zero-point constants.\footnotemark[6]
\footnotetext[6]{Limits were calculated using q$_{\circ}$ - 
2.5~log(5 $\times$ rms $\times$ $\sqrt{N_{\rm pix}}$ ), 
where N$_{\rm pix}$ and rms
are the number of pixels and variance within the 2$\arcsec$ diameter
aperture respectively, and q$_{\circ}$ is the zero-point constant.}
Typical magnitude limits are
27.0, 25.7, and 25.0 (Vega) in $B_W$, $R$, and $I$, respectively. 
These are listed in Table 2 for each source.  Figure 3 shows a histogram
of the integrated 20 cm continuum flux densities for the OIRS sample. 
Nearly 90$\%$ are sub-mJy radio sources (median F$_{\rm 20cm}^{\rm total}$
= 0.40 mJy), though five have flux densities greater than 1 mJy.  Source
$\#$362 possesses the peak flux density for the sample 
of nearly 7 mJy.  Thirty-one OIRSs were included within the MIPS survey area.
Only four of these - sources $\#$ 97, 176, 245, and 363 in Table 2 - were 
detected at 24 $\mu$m above 5 $\sigma$. We stack-averaged the remaining 
twenty-seven MIPS sub-images in an attempt
to detect fainter emission levels. No significant signal was detected,
and we set an upper-limit on the average infrared flux density for
the ensemble of F$_{\rm 24 \mu m}$ $<$ 60 $\mu$Jy (5 $\sigma$).

The small number of radio/infrared matches (13$\%$) is a significant
result, since from the surface densities of the 20 cm and 24 $\mu$m
sources, we estimate the probability of a random match to be less than
1$\%$.  We verified the lack of overlap between these two source populations
by plotting the OIRSs' positions on the MIPS 24 $\mu$m image and inspecting
the result visually. Two of these MIPS detected sources are shown in
the bottom two rows of Figure 2.  Eight OIRSs are situated in the survey
area covered by both Chandra and MIPS (0.08 deg$^{2}$).
None were detected at 24 $\mu$m or in 
any of the X-ray bands. This sets upper-limits of 0.3 mJy at 24 $\mu$m 
and 1.5 $\times$ 10$^{-16}$ erg s$^{-1}$ cm$^{-2}$ in the 0.5-2.0 keV 
X-ray band for this sub-sample.

Because of the variable sensitivity throughout the radio images, simply dividing 
the number of OIRSs by the survey area would lead to an under-estimate
of their true surface density ($\sigma_{\rm OIRS}$). This is
due to the fact that the fainter radio sources can only be detected near
the pointing centers where the sensitivity is the highest. We have
attempted a first-order correction to this effect by first weighting each OIRS
by the ratio of the total survey area to the area where it could have been
detected given its peak 20 cm flux density (Hopkins et al. 1999).
Doing so resulted in an average surface density of $\sigma_{\rm OIRS}$ =
104 $\pm$ 17 deg$^{-2}$, where the uncertainty represents Poisson counting
errors only.  There is a strong dependence on flux density. We divided
the OIRSs into three flux density bins, F$^{\rm peak}_{\rm 20 cm}$
$<$ 400 (17 sources), 400-600 (13 sources), and $>$ 600 (6 sources)
$\mu$Jy beam$^{-1}$, and calculated the
average surface density for each as above. We find that
$\sigma_{\rm OIRS}$($<$ 400 $\mu$Jy beam$^{-1}$) = 66
$\pm$ 12 deg$^{-2}$, $\sigma_{\rm OIRS}$(400-600 $\mu$Jy beam$^{-1}$) = 
26 $\pm$ 11 deg$^{-2}$, and $\sigma_{\rm OIRS}$($>$ 600 $\mu$Jy beam$^{-1}$) 
= 12 $\pm$ 5 deg$^{-2}$, i.e., the OIRS surface density increases with
decreasing 20 cm flux density.

\subsection{ Are The OIRSs Powered By Star Formation Or AGN? }

The logarithm of the infrared to radio continuum flux density ratio,
or {\em q}, is a useful parameter for distinguishing sources
powered by massive star formation from AGN, and especially radio-loud AGN.  
In Figure 4 we show the distribution of q using the MIPS 24 $\mu$m and 
20 cm radio continuum data in Table 2 for the OIRS sample
(i.e., q = log(F$_{\rm 24 \mu m}$/F$_{\rm 20 cm}$)).  The four OIRSs
detected by MIPS are represented by the shaded histogram
bins (-0.5 $\le$ q $\le$ 0.5).  For the twenty-seven other OIRSs
we show a histogram of upper-limits for q, using F$_{\rm 24 \mu m}$ = 
0.3 mJy in the calculation.
The small number of sources precludes a strong conclusion statistically, but
we note that three of the four OIRSs with MIPS detections have positive values 
of q, with a mean of 0.29.  Only source $\#$97 has
a negative q. On the other hand, of the 27 OIRSs not 
detected by MIPS, two-thirds (18 of 27) have negative q values. 
Since these are all upper-limits, it is likely that even a larger fraction
of these sources will have q $<$ 0.  

We first considered the origin of the OIRSs' emission
by comparing the values and upper-limits of q in Table 2
to those derived from local starbursts and AGN for a range of redshifts.
Using the most luminous galaxy (3$\times$10$^{12}$ L$_{\odot}$) SED from Lagache 
et al. (2002), we calculated the observer's frame 
F$_{\rm 850 \mu m}$/F$_{\rm 24 \mu m}$ ratio
for redshifts of 1, 2, and 3.  This in turn was used with the scaling
in Chary \& Elbaz (2001) to calculate the corresponding 20 cm flux densities.
Under these assumptions, starburst galaxies with 20 cm
flux densities of 200 $\mu$Jy should have 24 $\mu$m flux density of 2.8 mJy 
at z = 1, 2.4 mJy at z = 2, and 1.6 mJy at z = 3.  In short, starbursts detected 
in our radio survey would be easily detected in our 24 $\mu$m data.
These will in turn give rise to values of q equal to 1.15, 
1.08, and 0.90 for starbursts at these three redshifts.  
This is in agreement with Appleton et al. (2004), who found that for 
flux densities measured at 24 $\mu$m and 20 cm, starburst dominated systems 
have q = 0.8 $\pm$ 0.3 for z $\la$ 2. This range is represented in Figure 4 as the solid and
dashed vertical lines.  The values of q for starburst dominated systems are
substantially greater than the upper-limits shown,
and even appear significantly larger than the q's determined for the four
OIRSs detected by MIPS. Thus for virtually the entire OIRS sample
a clear enhancement in radio continuum emission over that expected for
a pure starburst is evident.  This indicates the presence of a dominant AGN
component.

A more detailed illustration with a broader range of SED type is presented 
in Figure 5, where q is plotted against the logarithm of 20 cm flux density. 
Four families of curves are shown, representing
(from top to bottom) (1) a normal Sc galaxy with a SFR of 3
M$_{\odot}$ yr$^{-1}$, (2) a high luminosity starburst/ULIRG with a
3000 M$_{\odot}$ yr$^{-1}$ SFR, (3) a type 2 Seyfert AGN (NGC 262), 
and (4) a QSO with 3C~273's spectral energy distribution. The late spiral,
ULIRG, and AGN/Seyfert SEDs were taken from the compilation of Xu et
al. (2001), while the 3C~273 SED was taken from the NED online database.
For the two star forming galaxy templates, we first calculated the rest-frame
20 cm luminosity (L$_{\rm 20 cm}$ in W Hz$^{-1}$) for a given star 
formation rate (in M$_{\odot}$ yr$^{-1}$) using the calibration given in
Equation 6 of Bell (2003). Using this value and the SED we derived
the observer-frame F$_{\rm 24 \mu m}$/F$_{\rm 20 cm}$ ratio as a function of
redshift from z = 0.1-7.0.  We next calculated the observer-frame 20 cm
flux density (F$_{\rm 20 cm}$, in $\mu$Jy) for these redshifts using
\begin{equation}
{\rm
L_{\rm 20 cm} = 1.26 
\times 10^{-35}~F_{\rm 20 cm}~d^{2}_{\rm L}~(1+z)^{\alpha-1}~~~(W~Hz^{-1}),  }
\end{equation}
where d$_{\rm L}$ is the luminosity distance in cm.
The radio spectral index $\alpha$ (where F$_{\rm \lambda}$ $\varpropto$
$\lambda^{\alpha}$) was derived from the SED templates over
the wavelength range 7 to 20 cm.  The starburst, normal, and AGN templates
gave similar average spectral indices (mean $\alpha$ = 0.65).
For the two AGN we considered 20 cm luminosities of 10$^{25}$ and 10$^{26}$ 
W Hz$^{-1}$, and calculated observer-frame F$_{\rm 24 \mu m}$/F$_{\rm 20 cm}$
ratios and 20 cm flux densities as above, also for the redshift range z = 0.1-7.0.
The resulting four sets of tracks are indexed by redshift from z = 0.1 to 7. 

The four MIPS detected OIRSs
are shown as unfilled squares in Figure 5.  For three ($\#$176, 245, 
\& 363), their positions are consistent with the SFR = 3000 M$_{\odot}$ yr$^{-1}$
ULIRG template over a z=2-3 redshift range, albeit somewhat displaced below 
the track. Note that increasing/decreasing the SFR for either the ULIRG or Sc galaxy
templates has the effect of moving the tracks to the right/left.  The vertical 
position is controlled solely by the shape of the SED and redshift.
It is therefore possible that these could be at z = 4-7 if of sufficient
luminosity.  The other MIPS detected OIRS ($\#$97) is situated well below 
the ULIRG track, with q = -0.24 $\pm$ 0.21.  The remaining twenty-seven OIRSs 
with q upper-limits are shown as filled circles with arrows.  Again, two-thirds
of them have negative values of q, which is inconsistent
with late spiral or ULIRG SEDs at any redshift. These sources show a clear
excess in radio emission relative to the infrared compared to a system powered by
star formation. The q upper-limits cannot be used to distinguish
between Seyfert and quasar templates in Figure 5 (both require q $\la$ -1 for
all z). However, it is clear that the bulk of the OIRSs are powered
primarily by an AGN, with many being radio-loud.

\subsection{ Are The OIXSs Starbursts Or AGN Powered Systems? }

X-ray sources with very faint optical counterparts, or no counterparts 
at all, are also found in deep surveys (e.g., Barger et
al. 2003).  Like the optically faint radio source population, 
these objects have been proposed to represent highly
obscured starburst galaxies or AGN at very high redshift, or possibly
even AGN which are over-luminous in X-rays compared to local
examples. To investigate the nature of these sources, we examined
the eleven X-ray sources in Table 1 of Wang et al. (2004) without
$B_W$, $R$, \& $I$ counterparts that were observed by MIPS at 24 $\mu$m.
The properties of these OIXSs, including source number 
(taken from Table 1 in Wang et al. 2004), J2000 coordinates, ``soft'' band X-ray 
fluxes or upper-limits, and optical limits are given in Table 3.  Also 
listed are the OIXSs' X-ray ``hardness ratio'' (HR), defined as  
HR = (H-S)/(H+S), where H and S represent the ``hard'' 2-7 keV
and ``soft'' 0.5-2 keV band counts, respectively.

No OIXSs were detected at 24 $\mu$m.  This has important
consequences for the origin of their X-ray luminosity. Weedman et al. 
(2004) used infrared and X-ray spectra from a sample of nine starburst regions 
in four merging systems and three obscured AGN to determine
empirically that the ratio IR/X (defined to be the 24 $\mu$m flux density
in mJy to the Chandra 0.5-2.0 keV flux in units of 10$^{-16}$ erg s$^{-1}$
cm$^{-2}$) has a lower bound of 0.2 for starbursts and
obscured AGN for z $\le$ 3.  Ratios of IR/X smaller than 0.2 indicate that 
the source is powered by a relatively unobscured AGN.  Note that we
should detect any ``soft'' band X-ray source in Wang et al. (2004)'s survey that
is powered by a starburst or heavily obscured AGN (i.e., IR/X $>$ 0.2) given
the Chandra and MIPS 24 $\mu$m sensitivities.
This empirical criterion can be used to classify the seven OIXSs in
our MIPS survey that were detected in the ``soft'' X-ray band.
Figure 6 shows a plot of IR/X versus the X-ray hardness ratio for these
sources. All seven have values of IR/X $<$ 0.2, which clearly indicates
the presence of a dominant unobscured AGN. We can not classify the
four OIXSs in Table 3 with ``soft'' band upper-limits. Nevertheless, at 
least 64$\%$ (7/11) of the OIXSs are powered by unobscured AGN.

The X-ray hardness ratio can be used to place rough constraints on the redshift
distribution of the OIXSs.  Wang et al. (2004) argued that the X-ray spectrum
will soften with increasing redshift, with HR $<$ 0.6 marking z$\ga$ 1 objects.
In support of this, Wang et al. (2004) determined photometric redshifts for five X-ray
sources detected at $I$ and {\em z'} bands but not in $R$. The three with 
z$_{\rm phot} \sim$ 1-2
have hardness ratios between -0.1 and 0.6, while the two ``softest'' sources
(HR $\la$ -0.5) have z$_{\rm phot} \sim$ 4.3. By this criteria,
nine out of the eleven OIXSs in our sample can have z $>$ 1 (HR $<$ 0.6). 
Four can possibly at z $\sim$ 4-5 (HR $<$ -0.5).

No OIXSs were detected by our radio survey, which sets a 20 cm flux density
limit of $\sim$200 $\mu$Jy for these sources.  Note that at these levels we
would have easily detected radio loud quasars with 3C~273-like
SEDs and 20 cm luminosities greater than 10$^{25}$ W Hz$^{-1}$
beyond redshifts of six (Figure 5), and most Seyfert
systems with similar 20 cm luminosities.  OIXSs thus appear to be largely
a population of radio-quiet AGN.

\subsection{Comparison Of Radio, Infrared, \& X-Ray Source Catalogs}

This study is concerned primarily with the optically invisible source
populations.  However, it is worthwhile considering the characteristics
of the radio, infrared, and X-ray populations as a whole.  Because of the
tight correlation between infrared and radio flux densities for
infrared-luminous galaxies (de Jong et al. 1985; Helou et al. 1985)
and because of the indications that the sub-mJy radio population corresponds
to faint, star-forming galaxies (e.g., Haarsma et al. 2001), a strong overlap
between the radio and MIPS survey samples might be expected.  Such an overlap
is implied by the results displayed from analogous work in 
the First Look Survey (FLS) area, for 
which Appleton et al. (2004) show a clear correlation 
between flux densities at 24 $\mu$m and at 20 cm (see their Figure 2).  
However, the sources plotted there are only those detected in both infrared 
and radio with redshifts less than $\sim$2, which does not provide 
a comparison of the entire radio and infrared samples in the FLS survey region.

Table 4 shows various source detection statistics for our radio, 
infrared, and X-ray survey regions.
First, we note that our radio survey does not go sufficiently deep to detect
the majority of MIPS sources. Within the 0.5 deg$^{2}$ VLA survey area 
there are 1405 MIPS detections, and 377 compact radio sources.  Only 125 are
detected in both wavebands. That is, 9 $\pm$ 1 $\%$ of the infrared sources
have radio counterparts and 33 $\pm$ 3 $\%$ of the radio sources have MIPS
counterparts with fluxes of at least 0.3 mJy.  This is not too surprising since
our radio observations were intended to detect the F$_{\rm 24 \mu m} \ge$ 0.75
mJy source population.  There are 216 such 24 $\mu$m sources in our
survey area. Only 33 $\pm$ 5 $\%$ of these have radio counterparts.
Similarly, 23 $\pm$ 3 $\%$ of the compact radio sources have infrared
counterparts $>$ 0.75 mJy at 24 $\mu$m.  
Sensitivity variations will of course introduce a bias against
detecting faint radio sources far away from the three pointing centers.  However,
we do not see a large effect.  If we restrict ourselves to regions where the radio
rms is less than 17 $\mu$Jy beam$^{-1}$, i.e., an area of 0.17 deg$^{2}$, we
find similar detection fractions: 32 $\pm$ 7 $\%$ (25/78) of the F$_{\rm 24 \mu~m} \ge$
0.75 mJy sources have radio counterparts, while 19 $\pm$ 4 $\%$ (25/130) of the 
compact radio sources MIPS counterparts with F$_{\rm 24 \mu~m} \ge$ 0.75 mJy.
This shows that greater 20 cm sensitivities are required to detect the bulk of the
infrared sources for which IRS spectra can be obtained in reasonable integration times.

When comparing the radio and infrared populations, it is therefore
important to realize that only a relatively small fraction of 
sources are detected in both bands.  This is illustrated in
Figure 7, where we show histograms of q for the 125 sources
detected at both 24 $\mu$m and 20 cm (thick solid line),
the 194 radio sources not detected at 24 $\mu$m (thin dashed line),
and the 1086 24 $\mu$m sources with no measured radio fluxes (thick
dashed line).  For the radio-only sample we calculated q 
upper-limits using F$_{\rm 24 \mu~m}$ = 0.3 mJy. 
For the infrared-only detected sample we calculated lower-limits for
q by setting F$_{\rm 20 cm}$ equal to that of a point source with a peak
flux density 5$\times$ the radio image rms at the infrared source's
position. 

The sources detected at both 24 $\mu$m and 20 cm show a distribution
of q that peaks at $\sim$ 0.5. This is in reasonable 
agreement with Appleton et al.'s (2004) result that q = 0.8 $\pm$ 0.3 
for the radio-infrared correlation at z $\la$ 2. The smaller range
of q evident in Figure 7 may be partly due to the detection of 
z $>$ 2 ULIRGS, which will have progressively smaller q with
increasing redshift (see the ULIRG track in Figure 5).
This is consistent with a population of galaxies
powered primarily by star formation, though an AGN contribution
is still possible.  Radio-loud AGN are certainly
apparent in the negative q tail. AGN also clearly dominate the population 
of radio sources with no MIPS detections in Figure 7.
However, the bulk of the infrared sources in the radio
survey field have only lower-limits for q, with a strong peak
between 0.5 and 1. While this is consistent with star formation
being the dominant power source, deeper 20 cm observations will be 
required to measure q and quantify the radio-infrared correlation 
for the majority of galaxies in the survey. Figure 7 suggests that
the radio and infrared selected populations do not overlap, i.e.,
that each represents populations of AGN and starburst dominated galaxies,
respectively. Higher sensitivity radio and infrared observations will
be needed to verify this.

An important related question is whether the OIRSs represent a population
significantly different from the radio sources with optical
counterparts, whose characteristics (e.g., starburst or AGN dominated)
can be determined through optical or near-infrared spectroscopy. There
is a qualitative indication that this might be so from the overall
difference in the infrared detection rate for radio sources in the
OIRS sample compared to the full radio sample.  We find that
33 $\pm$ 3$\%$ of the compact radio sources in the entire sample are detected in
the infrared to the 5 $\sigma$ sensitivity limit, whereas only 13 $\pm$ 7 $\%$ of the 
OIRS are detected.  Here the uncertainties represent Poisson statistics only. 
While this suggests a difference in the two radio populations, the small
number of sources in the OIRS sample prevents this from being a
statistically robust conclusion.

In the portion of the field covered by both Chandra and Spitzer surveys
($\sim$0.07 deg$^{2}$), there are 240 24 $\mu$m sources brighter than 0.3 mJy
(see Table 4), and 104 Chandra ``soft'' band sources
brighter than 1.5$\times$10$^{-16}$ ergs s$^{-1}$ cm$^{-2}$.  Only
18 sources are detected in both infrared and X-ray bands, for which
IR/X can be determined directly. A comparison of these sources is shown 
in Figure 8. For the eighteen sources detected in both bands (shaded histogram), 
nearly all (80$\%$) have IR/X $\le$ 0.2, indicating a relatively unobscured AGN.
For sources detected by Chandra but not MIPS, IR/X
upper-limits were determined as 0.3 mJy/(F$_{\rm 0.5-2.0 keV}$) with
the ``soft'' X-ray flux in units of 10$^{-16}$ erg s$^{-1}$ cm$^{-2}$.
For sources detected only at 24 $\mu$m, we calculated lower-limits to IR/X
using F$_{\rm 24 \mu m}$ in mJy divided by 1.5.  Figure 8 shows a clear 
separation between the sources detected only by MIPS or Chandra, with the 
former being identified with starburst or obscured AGN dominated systems and 
the latter with unobscured AGN. Quantitatively, 90$\%$ of the Chandra-only 
detections have IR/X $<$ 0.2. Similarly, 90$\%$ of the MIPS-only detections have 
IR/X $>$ 0.2.  These results apply to the full sample in the overlapping area, 
virtually all of which have optical identifications, and indicate that the great 
majority of X-ray sources are powered by AGN.

\section{Discussion}

\subsection{   Space Densities Of The OIRS And OIXS Populations  }

We of course have no direct measure of the redshifts for the OIRS
population. However the requirement that they be optically ``invisible'' likely
forces the majority to be at z $\ga$ 1.  
Supporting evidence for this comes from IRS spectroscopy of a sample of sixteen
optically faint and invisible MIPS sources in Bo\"otes with F$_{\rm 24 \mu m}$ 
$\ge$ 0.75 mJy in Houck et al. (2005).  For the ten optically invisible infrared
sources, the median spectroscopic redshift was 2.1, with only one object
having a redshift less than one, and that at z = 0.7 $\pm$ 0.1. 
We can calculate rough estimates of the OIRS space density by assuming 
that they  lie within certain broad redshift ranges.  By comparing 
these with measured or estimated space densities of other extragalactic
objects (see Table 5) we can gain additional clues into their nature. 
Arguments based on surface density comparisons are subject to
various complexities and assumptions (i.e., the OIRS redshift range) 
that are difficult to take into account. Nevertheless, such an approach is 
still informative.

The average surface 
density of the 27 OIRSs with no MIPS detections, sources whose emission is likely 
to be dominated by an AGN, is $\sigma_{\rm OIRS}$ = 80 $\pm$ 15 deg$^{-2}$.  
For our adopted flat $\Lambda$CDM cosmology and H$_{0}$, this
corresponds to a mean space density $\rho_{\rm OIRS}$ = (2.4 $\pm$ 0.5) 
$\times$ 10$^{-6}$ Mpc$^{-3}$ over a 1 $<$ z $<$ 5 redshift range. This is 
much smaller than the space density of current day massive ellipticals 
($\rho_{\rm elliptical}$ $\sim$ 10$^{-3}$ Mpc$^{-3}$, Marzke et al. 
1994), though comparable to that of local radio galaxies
($\rho_{\rm RG}$ $\sim$ 10$^{-6}$ Mpc$^{-3}$, 
Osterbrock 1989). 

This can also be compared with the
space density of bright (M$_{\rm B}$ $<$ -26) quasars over 0.4 $<$ z $<$ 5 
shown in Figure 3 of Fan et al. (2001), which combines results from
2dF (Boyle et al. 2000), Warren et al. (1994), and Schmidt et al. (1995),
as well as the Sloan Digital Sky Survey.  The OIRS space density is considerably 
larger than that of bright optically selected quasars, 
even at z $\sim$ 2.5 where the quasar distribution peaks 
($\rho_{\rm qso}$ $\sim$ 5 $\times$ 10$^{-7}$ Mpc$^{-3}$).  
The discrepancy is much larger at lower and
higher redshifts.  For example, from z = 3-5 the bright quasar space density
drops from $\sim$2 $\times$ 10$^{-7}$ to $\sim$1 $\times$ 10$^{-8}$
Mpc$^{-3}$, which is smaller than the OIRS space density by one and two
orders of magnitude, respectively. 

However, the OIRSs are almost certainly
less luminous optically than M$_{\rm B}$ = -26.
Assuming a power law form for the optical continuum of
f$_{\nu}$ $\varpropto$ $\nu$$^{-0.5}$, the $B_W$ upper limit of $\sim$26.5 
implies absolute magnitude upper limits of approximately M$_{\rm B}$ =
-18 to -23 for z=1-5. These values are less than or equal to M$_{\rm B}$ = -23, 
the traditional boundary between Seyfert galaxies and quasars.  The
cumulative luminosity function for optically selected quasars at z $>$ 3.6 in
Fan et al. (2001) is,
\begin{equation}
{\rm
 log\Phi(z,<M_{\rm B}) = (-6.91 \pm 0.19) - (0.48 \pm 0.15)(z-3) +
  (0.63 \pm 0.10)(M_{\rm B}+26)
}
\end{equation}
in units of Mpc$^{-3}$. This equation is strictly valid for M$_{\rm B}$ $<$ -25.6,
and employing it for intrinsically fainter objects requires that the
luminosity function's slope does not change significantly, which
we admit is an assumption.  Nevertheless,
using Equation 2 to calculate the space density of objects at
z = 5 with -23 $<$ M$_{\rm B}$ $<$ -22 gives a value of $\rho$ = 3 $\times$
10$^{-6}$ Mpc$^{-3}$, which is similar to the mean OIRS space density.
For the same range in M$_{\rm B}$, we derive space densities
of 3 $\times$ 10$^{-5}$ and 1 $\times$ 10$^{-5}$ Mpc$^{-3}$ at
redshifts of three and four, respectively. 
At a redshift of two, the 2dF cumulative luminosity function 
shown in Equation 4 of Fan et al. (2001) gives $\rho$ = 1 $\times$ 10$^{-6}$
Mpc$^{-3}$ for M$_{\rm B}$ $<$ -25.5.  For a range in M$_{\rm B}$
from -25 to -24 the extrapolated space density would
be $\sim$3 $\times$ 10$^{-6}$ Mpc$^{-3}$, which is also similar to
the mean $\rho$ for the OIRS population. Thus,
within the uncertainties accrued by extrapolating the bright high-z
quasar luminosity function to M$_{\rm B}$$\sim$-21, we conclude
that the averaged space density of OIRSs is consistent with 
that of lower luminosity AGN with redshifts 2 $<$ z $<$ 5.
Moreover, the distribution of q values for the OIRS population clearly
implies a strong radio-loud AGN component.

The average space density of sub-mm galaxies provides another
point of comparison.  These objects are thought to be the progenitors
of massive bulges that are powered primarily by heavily obscured star 
formation. Taking the sources detected at 850 $\mu$m by Barger et al. 
(1999) with fluxes greater than 2.2 mJy and assuming they lie 
between 1 $<$ z $<$ 3 would result in a space density of 
$\rho_{\rm 850 \mu m}$ = 7 $\times$ 
10$^{-5}$ Mpc$^{-3}$. This would be nearly sixteen times larger than the 
average OIRS space density over the same redshift interval, which we
determine to be $\rho_{\rm OIRS}$ = (4.1 $\pm$ 0.8) $\times$ 10$^{-6}$ 
Mpc$^{-3}$ if all OIRSs are within this interval.  
The sub-mm galaxies have a substantially larger space 
density.  The discrepancy becomes even worse if
one extrapolates the 850 $\mu$m source counts to fainter sources in
order to account for the diffuse sub-mm background emission, in
which case the average space density of sub-mm sources becomes
comparable to the space density of present-day elliptical galaxies.  
Such a large difference in space density
suggests that the OIRSs are not the objects that make up the sub-mm
source population.

The average surface density of the eleven OIXSs is 157 $\pm$ 47 deg$^{-2}$. 
While formally $\sim$50$\%$ larger than that of the OIRS population 
determined in $\S$4.1, $\sigma_{\rm OIXS} \approx$ $\sigma_{\rm OIRS}$ given the
uncertainties.  In $\S$4.3 we argued that nine out of eleven are likely to have 
1 $<$ z $<$ 5 based on their X-ray hardness ratios.  Distributing these 
uniformly over this redshift range gives an average space density $\rho_{\rm OIXS}$
= (3.9 $\pm$ 1.3) $\times$ 10$^{-6}$ Mpc$^{-3}$, again comparable to
$\rho_{\rm OIRS}$.  Thus the conclusions reached
above for the OIRS population holds for OIXSs as well: $\rho_{\rm OIXS}$ is
larger than the space density of local radio galaxies and luminous
optically selected 1 $<$ z $<$ 5 quasars, but consistent with
the space density of lower luminosity AGN (M$_{\rm B}$ $>$ -23)
extrapolated from the quasar luminosity function of Fan et al. (2001).
If four of the OIXSs are at z $\sim$ 4-5 as is suggested by the HR 
values in Table 3 and Figure 6, then the OIXS space density over 1 $<$ z $<$ 3 
can not be larger than (3.3 $\pm$ 1.5) $\times$ 10$^{-6}$ Mpc$^{-3}$,
which is more than an order of magnitude smaller than the mean space density of
Barger et al.'s (1999) F$_{\rm 850 \mu m}$ $>$ 2.2 mJy sub-mm sources. 
OIXSs do not appear to be the sources making up the sub-mm galaxy population.  

While additional and independent observational evidence is clearly required to
place these conclusions on a more solid footing, the estimated space
densities of both OIRSs and OIXSs are consistent with lower luminosity AGN
hosts (M$_{\rm B}$ $\ga$ -23) at z $\sim$ 1-5, suggesting that we are seeing
the fainter end of the high redshift quasar luminosity function.

\subsection{Do The Optically Invisible Radio And X-Ray Sources Represent 
            Different Populations Of AGN?}

We concluded in $\S$4.2 that the OIRSs are predominately AGN because of
their 24 $\mu$m to 20 cm flux ratios, and we concluded in $\S$4.3 that
the OIXSs are likewise primarily AGN because of their low infrared to
X-ray flux ratios.  It is worth noting that these two independently
derived samples of optically invisible AGN do not overlap. In the common
radio and X-ray survey region (Table 4) there are 10 OIRSs, none of which are
detected in ``soft'' or ``hard'' X-ray bands by Wang et al. (2004).
Similarly, there are 14 OIXSs within our VLA survey.  None of these
are detected at 20 cm. 

We can ask whether this result is consistent with expectations from
previous deep parallel surveys at radio and X-ray wavelengths.  
The most relevant study is that of Bauer et al. (2003) of the
Chandra Deep Field North, which compared radio and X-ray sources at flux
levels comparable to our sample.  They used a variety of spectroscopic 
indicators, including optical and X-ray spectra to identify AGN. 
Their results showed that such AGN covered a wide range of X-ray to
radio flux ratio. Specifically, they found (their Figure 1) that AGN could
exist with 0.002 $<$ F$_{\rm X-ray}$/F$_{\rm 20 cm}$ $<$ 2.5,
where F$_{\rm X-ray}$ is the Chandra full-band flux in units of
10$^{-16}$ ergs s$^{-1}$ cm$^{-2}$ and F$_{\rm 20 cm}$ is in units of
$\mu$Jy. The upper bound could even exceed 2.5 as this value derives
from the upper limit of 40 $\mu$Jy for the radio sources not detected
in X-rays.  From this, given that the faintest full-band X-ray flux in 
Wang et al. (2004)'s survey is 5 $\times$ 10$^{-16}$ erg s$^{-1}$ cm$^{-2}$, 
an AGN could have a 20 cm flux density as high as $\sim$2 mJy and still 
not be detected in X-rays. Of our OIRS sample, only a source like 
$\#$362 with F$_{\rm 20 cm}$ = 6.7 mJy would have been easily 
detected in X-rays.  The OIRSs' median 20 cm flux density is 400 $\mu$Jy.
For such a source to have a full-band X-ray flux less than 5 $\times$
10$^{-16}$ erg s$^{-1}$ cm$^{-2}$ would imply F$_{\rm
X-ray}$/F$_{\rm 20 cm}$ $<$ 0.013, which is well within the possible
range found by Bauer et al. (2003).

Conversely, the median OIXS has full band X-ray flux of
5 $\times$ 10$^{-15}$ ergs s$^{-1}$ cm$^{-2}$.  In order not to be
detected in our radio survey, it would have to be fainter than 100
$\mu$Jy, or have F$_{\rm X-ray}$/F$_{\rm 20 cm}$ $>$ 0.5, which is
also well within the observed range of this ratio.  We conclude,
therefore, that the lack of overlap between the radio and X-ray
samples of optically invisible AGN is consistent with the wide
range of F$_{\rm X-ray}$/F$_{\rm 20 cm}$ that has been established
among AGN.

\section{Summary}

We have combined a Spitzer/MIPS survey of the NDWFS
Bo\"otes field at 24 $\mu$m with an A-array
VLA 20 cm survey of a 0.5 deg$^{2}$ subregion
and a 172 ks Chandra observation to investigate the nature of
optically ``invisible'' radio and X-ray sources.  These we define
to be compact radio and X-ray sources without visible counterparts
in the NDWFS $B_W$, $R$, \& $I$ images. It has been proposed that 
these objects represent populations of high redshift galaxies harboring 
heavily obscured starbursts or AGN.
From the VLA and Chandra surveys we identify 31 OIRSs and 12 OIXSs
within the area surveyed by MIPS at 24 $\mu$m, out of total radio and
X-ray source populations of 377 and 168, respectively.  Only four of
the 31 OIRSs and none of the OIXSs are detected by MIPS at 24 $\mu$m.  
We compared the 20 cm and 24 $\mu$m emission properties of the OIRSs
observed with MIPS
with those expected from late spiral, ULIRG, Seyfert, and quasar
SEDs over a wide redshift range.  We conclude that the OIRSs are
primarily a population of galaxies powered by AGN rather 
than dust enshrouded starbursts, and that they likely lie at z $>$ 1.
Likewise, of the OIXSs with measured ``soft'' X-ray fluxes, all have 
24 $\mu$m to 0.5-2 keV band flux ratio limits consistent with a dominant 
and relatively unobscured AGN. The X-ray properties of OIXSs suggest
that most are likely at z $>$ 1, with several possibly at z $\sim$ 4.
Given the wide range in X-ray and radio properties of AGN, the fact that
the OIRSs and OIXSs have no object in common does not rule out their being 
from the same AGN source population.

Assuming the OIRSs populate the redshift range 1 $<$ z $<$ 5, their average
space density is (2.4 $\pm$ 0.5) $\times$ 10$^{-6}$ Mpc$^{-3}$.  This
is consistent with the space densities derived using the cumulative
quasar luminosity function from Fan et al. (2001) over the redshift
range z = 2-5 for -23 $<$ M$_{\rm B}$ $<$ -22, i.e., AGN with blue
luminosities comparable to current epoch Seyfert galaxies.
The space density of 850 $\mu$m sources 
is much larger than this, suggesting that they are a fundamentally 
different class of object.  Similarly, the OIXSs have an average space 
density of (3.8 $\pm$ 1.2) $\times$ 10$^{-6}$
Mpc$^{-3}$ if they exist uniformly throughout the redshift range
1 $<$ z $<$ 5, which is similar to the OIRSs'. This is also
consistent with the space density of lower luminosity AGN 
extrapolated from Fan et al. (2001) for this redshift range.

\acknowledgments
We would like to thank Keven Xu for access to his library of galaxy spectral 
energy distributions, and both Robert Becker and David Helfand for valuable
discussions.  We also wish to thank the referee for suggestions that led
to improvements in the manuscript. This work is based in part 
on observations made with the Spitzer Space Telescope, which is operated by the 
Jet Propulsion Laboratory, California 
Institute of Technology under NASA contract 1407. Support for this work 
was provided by NASA through Contract Number 1257184 issued by JPL/Caltech.
The National Radio Astronomy Observatory is a facility of the National
Science Foundation operated under cooperative agreement by Associated
Universities, Inc.  This publication made use of 
photographic data from the National Geographic Society -- Palomar
Observatory Sky Survey (NGS-POSS) obtained using the Oschin Telescope on
Palomar Mountain.  The NGS-POSS was funded by a grant from the National 
Geographic Society to the California Institute of Technology.  The      
plates were processed into the present compressed digital form with     
their permission.  The Digitized Sky Survey was produced at the Space   
Telescope Science Institute under US Government grant NAG W-2166.
This research has also made use of the NASA/IPAC Extragalactic Database (NED)
which is operated by the Jet Propulsion Laboratory, California Institute
of Technology, under contract with the National Aeronautics and Space
Administration.


\clearpage
\begin{table}
\tablenum{1}
\caption{VLA Observing Parameters}
\begin{tabular}{lccc}
\tableline\tableline
Observing Dates:                 & &26, 27 June $\&$ 27 July 2003 &\\
Array Configuration:             & &A&\\
Min./Max. Baseline (km):           & &0.68/36.4 &\\
Number of Antennas:              & &27&\\
Primary Beam FWHM (arcmin):     & &31.5 &\\
IF Frequencies (GHz):           & & 1.3649 $\&$ 1.4351 &\\
Number of Channels per IF:               &&7&\\
Channel Separation (MHz):      &&3.125&\\
Effective Bandwidth (MHz):        &&43.750&\\
Phase $\&$ Gain Calibrator:            &&1416+347&\\
Flux $\&$ Bandpass Calibrator:         &&3C~48&\\
Synthesized Beam FWHM (arcsec):\tablenotemark{a}   &&1.4&\\

\\
\tableline
Field Centers    &    North &       East&        South \\
\tableline\tableline
R.A.~(J2000): &14:25:57.00   &  14:27:27.00   & 14:25:57.00\\
Dec.~(J2000): &35:32:00.0~    &  35:17:12.0~    & 35:02:24.0~ \\
Time on Source (hrs):  &    5.5   &          4.9    &          5.4\\
Map r.m.s. ($\mu$Jy/beam):  & 14.8 & 15.5& 15.6\\

\tablenotetext{a} {Robust weighting.}
\end{tabular}
\end{table}

\newpage
\clearpage
\begin{deluxetable}{ccccccccccc}
\rotate
\tablenum{2}
\tabletypesize{\footnotesize}
\tablecaption{Optically Invisible Radio Sources In the Bo\"otes Sub-Region}
\tablewidth{0pc}
\tablehead{
   \colhead{Radio} & \colhead{Source}  &  \colhead{F$_{\rm 20 cm}$\tablenotemark{(a)}} & 
   \colhead{F$_{\rm 20 cm}^{\rm peak}$} & \colhead{a $\times$ b} & \colhead{$\phi$} & 
   \colhead{F$_{\rm 24 \mu m}$\tablenotemark{(b)}} & \colhead{q\tablenotemark{(c)}}  & 
$B_W$\tablenotemark{(d)} & $R$\tablenotemark{(d)} &$I$\tablenotemark{(d)} \\
   \colhead{ID\#} &   & \colhead{($\mu$Jy)} & \colhead{($\mu$Jy)} & \colhead{(arcsec)} & 
   \colhead{(deg.)} & \colhead{($\mu$Jy)} &  & & & }
\startdata
   9 & J142822.75+351849.8 & ~511 $\pm$ 62 & ~128 &3.5 $\times$ 2.7 &~70 &\nodata & $<$-0.26 &$>$27.2 &$>$25.3& $>$25.1\\
  19 & J142813.97+351136.6 & ~565 $\pm$ 36 & ~378 &1.9 $\times$ 1.5 &126 &\nodata & $<$-0.30 & $>$27.1& $>$25.2& $>$24.5\\
  22 & J142810.60+350659.2 & ~308 $\pm$ 42 & ~125 &1.7 $\times$ 1.5 &128 &\nodata & $<$-0.04 & $>$26.9 & $>$24.9& $>$24.3\\
  49 & J142744.94+352616.5 & ~565 $\pm$ 32 & ~378 &1.9 $\times$ 1.5 &126 &\nodata & $<$-0.30 & $>$27.2& $>$25.3& $>$25.0\\
  52 & J142740.39+350117.5 & ~240 $\pm$ 38 & ~148 &2.2 $\times$ 1.4 &137 &\nodata & $<$~0.07 & $>$26.9 &$>$25.3& $>$24.2\\
  79 & J142713.75+352134.5 & ~948 $\pm$ 37 & ~602 &2.1 $\times$ 1.5 &137 &\nodata & $<$-0.53 & $>$26.8 &$>$25.2& $>$25.3\\
  92 & J142702.54+351215.2 & ~239 $\pm$ 34 & ~210 &1.5 $\times$ 1.4 &~49 &\nodata & $<$~0.07 & $>$27.1&$>$25.6& $>$24.7\\
  97 & J142701.06+351949.3 & ~396 $\pm$ 32 & ~271 &1.7 $\times$ 1.6 &~69 &234 $\pm$ 44 & ~-0.24 $\pm$ 0.21 & $>$27.2 &$>$25.8& $>$25.2\\
 110 & J142652.98+353351.1 & ~177 $\pm$ 30 & ~162 &1.8 $\times$ 1.3 &129 &\nodata & $<$~0.20 & $>$27.1 &$>$26.1 & $>$25.5\\
 114 & J142650.00+350711.0 & ~372 $\pm$ 47 & ~164 &2.4 $\times$ 1.8 &~40 &\nodata & $<$-0.12 &$>$26.8&$>$25.5&$>$25.0\\
 156 & J142632.14+353614.2 & ~799 $\pm$ 59 & ~262 &2.9 $\times$ 2.0 &164 &\nodata & $<$-0.46 & $>$27.0 & $>$26.0& $>$25.4\\
 176 & J142624.94+350614.6 & ~249 $\pm$ 35 & ~171 &1.9 $\times$ 1.5 &~14 &462 $\pm$ 44 &~~0.27 $\pm$ 0.17 &$>$27.0 &$>$25.5&$>$24.7\\
 182 & J142623.40+345821.1 & ~554 $\pm$ 39 & ~352 &2.1 $\times$ 1.6 &100 &\nodata & $<$-0.30 &$>$27.0& $>$25.7&$>$24.7\\
 185 & J142621.92+353114.7 & ~327 $\pm$ 35 & ~230 &1.6 $\times$ 1.6 &131 &\nodata & $<$-0.07 & $>$27.0 &$>$26.0 & $>$25.2\\
 208 & J142613.81+353154.4 & ~189 $\pm$ 33 & ~145 &1.6 $\times$ 1.5 &~76 &\nodata & $<$~0.17 & $>$27.0 & $>$26.0& $>$25.3\\
 209 & J142613.53+352810.9 & ~165 $\pm$ 31 & ~~77 &2.0 $\times$ 1.4 &~12 &\nodata & $<$~0.23 & $>$27.0 & $>$25.9 & $>$25.3\\
 232 & J142605.13+350604.2 & ~512 $\pm$ 18 & ~404 &1.6 $\times$ 1.5 &~88 &\nodata & $<$-0.26 &$>$26.9&$>$25.1&$>$24.7 \\
 245 & J142602.35+350907.9 & ~244 $\pm$ 29 & ~230 &1.6 $\times$ 1.3 &~~2 &440 $\pm$ 52 &~~0.26 $\pm$ 0.17 &$>$26.9&$>$25.2&$>$24.8\\
 278 & J142551.58+351543.2 & ~445 $\pm$ 70 & ~118 &2.6 $\times$ 2.3 &178 &\nodata & $<$-0.20 & $>$27.1 &$>$25.8 & $>$25.6\\
 282 & J142550.49+352935.4 & ~155 $\pm$ 30 & ~~77 &2.3 $\times$ 1.5 &~19 &\nodata & $<$~0.26 & $>$27.2 &$>$26.0& $>$25.3\\
 305 & J142544.84+351702.2 & ~777 $\pm$ 83 & ~165 &3.4 $\times$ 2.6 &~77 &\nodata & $<$-0.44 & $>$26.9 &$>$26.2& $>$25.2\\
 313 & J142543.00+353049.2 & 1165 $\pm$ 71 & ~296 &3.0 $\times$ 2.6 &~11 &\nodata & $<$-0.62 & $>$27.1 &$>$25.8& $>$25.2\\
 323 & J142537.12+345229.2 & ~181 $\pm$ 35 & ~126 &1.7 $\times$ 1.3 &~20 &\nodata & $<$~0.19 & $>$27.1 &$>$25.5&$>$25.0 \\
 346 & J142528.98+352824.8 & ~367 $\pm$ 36 & ~205 &2.0 $\times$ 1.7 &~21 &\nodata & $<$-0.12 & $>$27.0 &$>$26.1& $>$25.4\\
 349 & J142527.24+352649.7 & ~241 $\pm$ 47 & ~108 &2.5 $\times$ 1.7 &~80 &\nodata & $<$~0.07 & $>$27.1 &$>$25.9& $>$25.1\\
 362 & J142525.04+344913.2 & 6738 $\pm$ 69 & 1806 &3.0 $\times$ 2.3 &~44 &\nodata & $<$-1.38 &$>$27.0&$>$25.8&$>$24.9\\
 363 & J142524.84+352554.1 & ~145 $\pm$ 38 & ~~90 &2.1 $\times$ 1.5 &~38 &328 $\pm$ 52 &~~0.35 $\pm$ 0.31 & $>$27.2 &$>$25.9& $>$25.3\\
 375 & J142519.91+352417.8 & ~847 $\pm$ 50 & ~346 &2.5 $\times$ 1.9 &~42 &\nodata & $<$-0.48 & $>$26.9 &$>$26.0& $>$25.4\\
 380 & J142516.19+350248.1 & 1906 $\pm$ 40 & 1070 &2.0 $\times$ 1.7 &~79 &\nodata & $<$-3.83 & $>$26.9 &$>$25.5&$>$24.9\\
 388 & J142508.28+353901.9 & ~180 $\pm$ 34 & ~111 &2.3 $\times$ 1.3 &~56 &\nodata & $<$~0.19 & $>$27.1 &$>$26.0& $>$25.3\\
 389 & J142507.78+354210.4 & ~302 $\pm$ 21 & ~164 &2.1 $\times$ 1.7 &116 &\tablenotemark{(e)} & \nodata  & $>$26.9 &$>$26.1& $>$25.3\\
 393 & J142506.41+353813.4 & ~171 $\pm$ 14 & ~112 &2.2 $\times$ 1.4 &125 &\tablenotemark{(e)} & \nodata  & $>$27.0 &$>$25.8& $>$25.2\\
 410 & J142457.20+351620.5 & ~875 $\pm$ 75 & ~307 &2.9 $\times$ 1.9 &151 &\nodata & $<$-0.49 & $>$27.0 &$>$25.8& $>$25.3\\
 430 & J142445.36+353417.1 & 1410 $\pm$ 62 & ~423 &3.3 $\times$ 1.9 &102 &\tablenotemark{(e)} & \nodata  & $>$27.0 &$>$25.9& $>$25.2\\
 434 & J142444.01+351227.1 & ~588 $\pm$ 47 & ~262 &3.1 $\times$ 1.4 &~75 &\tablenotemark{(e)} & \nodata   &$>$27.0&$>$25.5&$>$24.7\\
 441 & J142435.59+351046.2 & 1486 $\pm$ 59 & ~484 &2.7 $\times$ 2.2 &~11 &\tablenotemark{(e)} & \nodata   &$>$26.5&$>$25.1&$>$24.4\\
\enddata
\tablenotetext{(a)}{Integrated 20 cm flux in $\mu$Jy.}
\tablenotetext{(b)}{Integrated 24 $\mu$m fluxes (or non-detection) at radio source position.}
\tablenotetext{(c)}{ q = log(F$_{\rm 24 \mu m}$/F$_{\rm 20 cm}$).}
\tablenotetext{(d)}{$B_W$, $R$, \& $I$ limits (5 $\sigma$) calculated in 2\arcsec~diameter apertures
                    centered on the OIRS positions. See $\S$4.1.}
\tablenotetext{(e)}{These OIRSs are outside the MIPS 24 $\mu$m survey area.}
\end{deluxetable}

\newpage
\clearpage
\begin{deluxetable}{cccccccc}
\rotate
\tablenum{3}
\tabletypesize{\footnotesize}
\tablecaption{Optically Invisible X-Ray Sources}
\tablewidth{0pc}
\tablehead{
   \colhead{X-Ray\tablenotemark{(a)}} & \colhead{Source} & \colhead{F$_{\rm 0.5-2 keV}$\tablenotemark{(b)}} & 
   \colhead{HR\tablenotemark{(c)}} & \colhead{IR/X\tablenotemark{(d)}} & \colhead{$B_W$\tablenotemark{(e)}} & 
   \colhead{$R$\tablenotemark{(e)}} & \colhead{$I$\tablenotemark{(e)}} \\
   \colhead{ID \#} & \colhead{} & \colhead{ } & \colhead{} & \colhead{}  & \colhead{ } & \colhead{} & \colhead{}  }
\startdata
    9  &  J142539.55+353357.9 &   ~6.9  &  -0.43$^{\rm +0.19}_{\rm -0.17}$ & $<$0.04 & $>$26.8& $>$25.9 & $>$25.3\\
   15  &  J142530.71+353911.3 &   11.9  &  ~0.12$^{\rm +0.11}_{\rm -0.11}$ & $<$0.03 & $>$27.1& $>$25.8 & $>$25.4\\
   31  &  J142555.40+353650.4 &   ~3.3  &  -0.56$^{\rm +0.33}_{\rm -0.25}$ & $<$0.08 & $>$26.9& $>$25.8 & $>$25.3\\
   51  &  J142546.33+353349.4 &   $<$1.5\tablenotemark{(f)}  & $>$0.56 & \nodata  & $>$27.0& $>$26.0 & $>$25.3\\
   69  &  J142531.17+353921.6 &   $<$1.5\tablenotemark{(f)}  & $>$0.73 & \nodata  & $>$26.9& $>$26.2 & $>$25.3\\
   70  &  J142530.63+353420.3 &   ~9.4  &  -0.42$^{\rm +0.16}_{\rm -0.14}$ & $<$0.03 & $>$27.0& $>$25.9 & $>$25.3\\
   73  &  J142526.68+353140.8 &   ~7.0  &  -0.13$^{\rm +0.19}_{\rm -0.18}$ & $<$0.04 & $>$27.0& $>$26.0 & $>$25.3\\
   96  &  J142558.13+353216.1 &   ~3.8  &  -0.33$^{\rm +0.30}_{\rm -0.26}$ & $<$0.08 & $>$26.9& $>$25.8 & $>$25.3\\
  103  &  J142544.20+354018.4 &   $<$1.5\tablenotemark{(f)}  & $>$0.86  & \nodata & $>$26.9& $>$25.9 & $>$25.4\\
  112  &  J142527.59+354012.1 &   ~3.2  &  ~0.10$^{\rm +0.25}_{\rm -0.26}$ & $<$0.09 & $>$27.1& $>$25.8 & $>$25.2\\
  120  &  J142522.40+353517.2 &   $<$1.5\tablenotemark{(f)}  & $>$0.40  & \nodata & $>$26.9& $>$25.8 & $>$25.2\\
\enddata
\tablenotetext{(a)}{X-ray source number from Table 1 in Wang et al. (2004).}
\tablenotetext{(b)}{Integrated ``Soft'' (0.5-2.0 keV) band flux in units of 
                    10$^{-16}$ erg s$^{-1}$ cm$^{-2}$ from Wang et al. (2004).}
\tablenotetext{(c)}{X-ray ``hardness'' radio from Wang et al. (2004). HR = (H - S)/(H + S), where H and S
are the ``hard'' (2-7 keV) and ``soft'' (0.5-2 keV) band counts.}

\tablenotetext{(d)}{The 24 $\mu$m to soft X-ray flux radio, in units of 
 mJy/(10$^{-16}$ erg s$^{-1}$ cm$^{-2}$).}
\tablenotetext{(e)}{$B_W$, $R$, \& $I$ limits (5 $\sigma$) calculated in 2\arcsec ~diameter apertures
                    centered on the OIXS positions as in $\S$4.1.}
\tablenotetext{(f)} {OIXSs $\#$51, 69, 103 $\&$ 120 were only detected in the Chandra ``hard'' or ``total''
                     bands, and not the 0.5-2 keV ``soft'' band. Their positional accuracies are all $<$ 1\arcsec.}

\end{deluxetable}

\newpage
\clearpage
\begin{table}
\tablenum{4}
\caption{Source Detection Statistics In Radio and X-ray Surveys}
\begin{tabular}{lccccc}
\\
\tableline
 {\bf Radio Survey}    & Total Number & $\sigma$\tablenotemark{(a)} & $\%$MIPS\tablenotemark{(b)} & 
    $\%$Radio\tablenotemark{(c)} &   \\
\tableline
MIPS Sources ($>$ 0.3 mJy) &  1405       &          2810 $\pm$ 75       &       \nodata      &       9          & \\
Compact Radio Sources&  ~377       &          ~~754 $\pm$ 39       &       33           &    \nodata       & \\
OIRSs         &  ~~36       & ~~104 $\pm$ 17\tablenotemark{(d)}  &  13\tablenotemark{(e)}   &    \nodata       & \\
\\
\tableline
 {\bf X-Ray Survey} & Total Number & $\sigma$\tablenotemark{(a)} & $\%$MIPS & $\%$Radio & $\%$X-ray\tablenotemark{(f)} \\
\tableline
X-Ray Sources &  104\tablenotemark{(g)} &  1486 $\pm$ 146         &     17               &  11      &  \nodata   \\
MIPS Sources ($>$ 0.3 mJy) &  240        &  3429 $\pm$ 221         &     \nodata          &  11      &  ~8        \\
Compact Radio Sources&  ~73        &       1043 $\pm$ 123         &     36               & \nodata  &  14       \\
OIXSs         &  ~11        &      ~157 $\pm$ ~47         &     ~0               &  ~0      &  \nodata  \\
OIRSs         &  ~~8        &      ~114 $\pm$ ~40         &     13               & \nodata  &  ~0       \\
\tablenotetext{(a)}{Average surface density in deg$^{-2}$.}
\tablenotetext{(b)}{Percent of sources detected at 24 $\mu$m by MIPS.}
\tablenotetext{(c)}{Percent of sources detected at 20 cm.}
\tablenotetext{(d)}{Surface density for all 36 OIRSs, including those outside the MIPS survey area.}
\tablenotetext{(e)}{Percentage of the 31 OIRSs observed by MIPS that were detected at 24 $\mu$m.}
\tablenotetext{(f)}{Percent of sources detected in Chandra 0.5-2.0 keV band by Wang et al. (2004).}
\tablenotetext{(g)}{Number of X-ray sources observed by MIPS at 24 $\mu$m.}

\end{tabular}
\end{table}

\newpage
\clearpage
\begin{deluxetable}{lccl}
\tablenum{5}
\tabletypesize{\footnotesize}
\tablecaption{Average Space Density Comparison}
\tablewidth{0pc}
\tablehead{
   \colhead{Objects} & \colhead{ $\rho$} & \colhead{Redshift}   & \colhead{Notes} \\
   \colhead{ }       & \colhead{ (10$^{-6}$ Mpc$^{3}$) } & \colhead{} & \colhead{}     
}
\startdata
OIRS   		& 3.1 $\pm$ 0.5 & 1 $<$ z $<$ 5 & Calculated using all 36 OIRSs in the VLA survey.\\
                & 2.4 $\pm$ 0.5 & 1 $<$ z $<$ 5 & Calculated using the 27 OIRSs not detected by MIPS.\\
                & 4.1 $\pm$ 0.8 & 1 $<$ z $<$ 3 &                             \\
                &&& \\
OIXS 		& 4.7 $\pm$ 1.4 & 1 $<$ z $<$ 5 & Calculated using all 11 OIXSs in Chandra survey. \\
                & 3.9 $\pm$ 1.3 & 1 $<$ z $<$ 5 & Calculated using the 9 OIXSs with HR $<$ 0.6. \\
                & 3.3 $\pm$ 1.5 & 1 $<$ z $<$ 3 & Calculated using the 5  OIXSs with -0.5 $<$ HR $<$ 0.6.\\
                &&& \\
Ellipticals     & 10$^{3}$      & $\sim$0   & Local massive ellipticals (Marzke et al. 1994).\\
                &&& \\
Radio Galaxies  & $\sim$1       & $\sim$0   & Local radio galaxies (Osterbrock 1989).\\
                &&& \\
Sub-mm Sources  & 70            & 1 $<$ z $<$ 3 & F$_{\rm 850 \mu m}$$>$ 2.2 mJy (Barger et al. 1999).\\
                &&& \\
Bright QSOs     & 0.01          &  z = 1      & M$_{\rm B}$ $<$ -26 QSOs. From Figure 3 in Fan et al. (2001).\\
                & 0.5-0.01      & 2 $<$ z $<$ 5 & M$_{\rm B}$ $<$ -26 QSOs. From Figure 3 in Fan et al. (2001).\\
                &&& \\
Faint QSOs      & 30-3          & 3 $<$ z $<$ 5 & -23 $<$ M$_{\rm B}$ $<$ -22 QSOs. Extrapolated from Fan et al. (2001).\\
\enddata
\end{deluxetable}


\clearpage
\begin{figure}
\figurenum{1}
\plotone{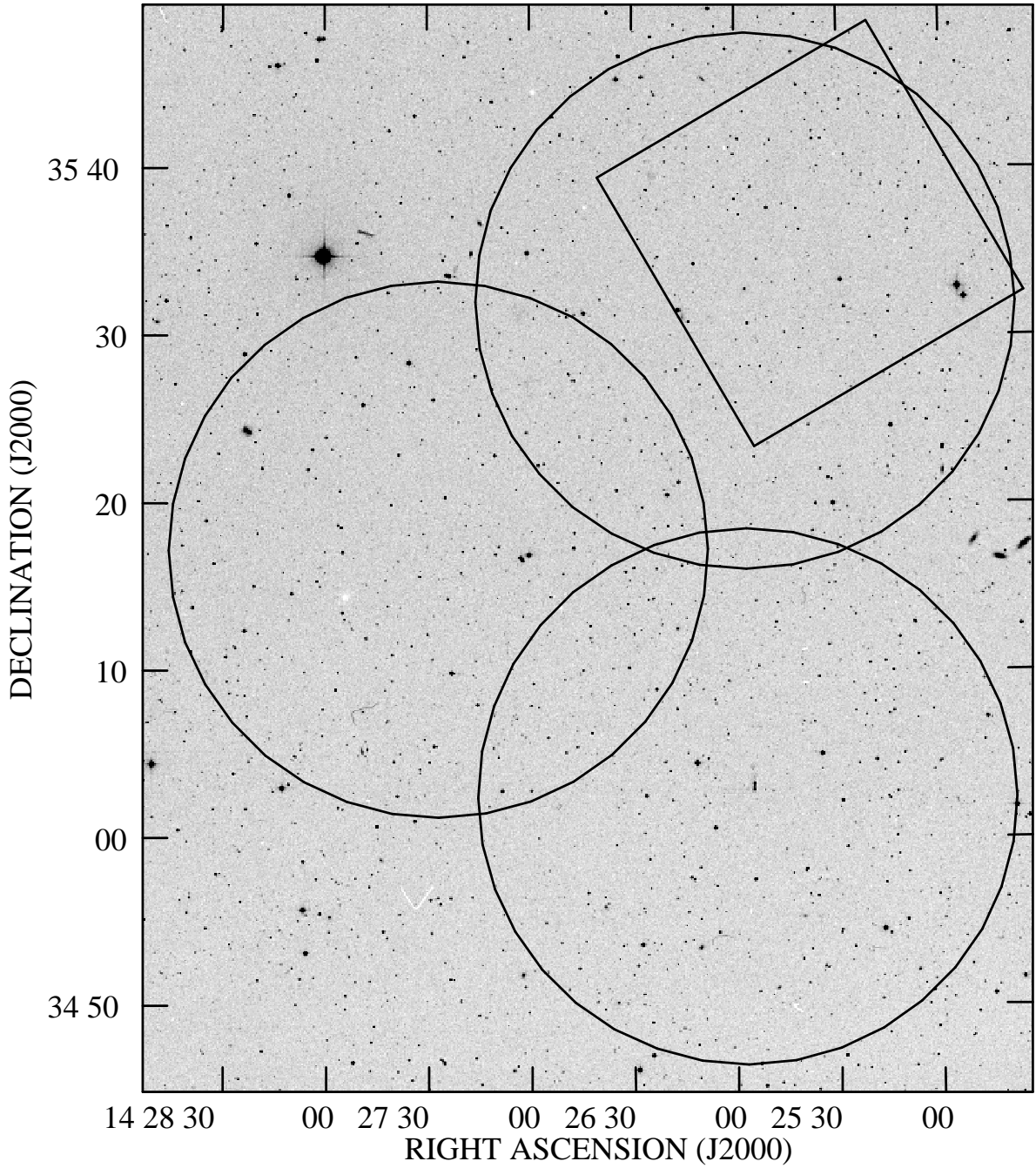}
\caption{The survey fields.  The circles represent the VLA
primary beam FWHM (32$'$) for the three array pointings (North, East,
 \& South), while the square represents the region observed with Chandra 
by Wang et al. (2004).  These are superposed on a DPOSS grey-scale image.}
\end{figure}

\clearpage
\begin{figure}
\figurenum{2}
\plotone{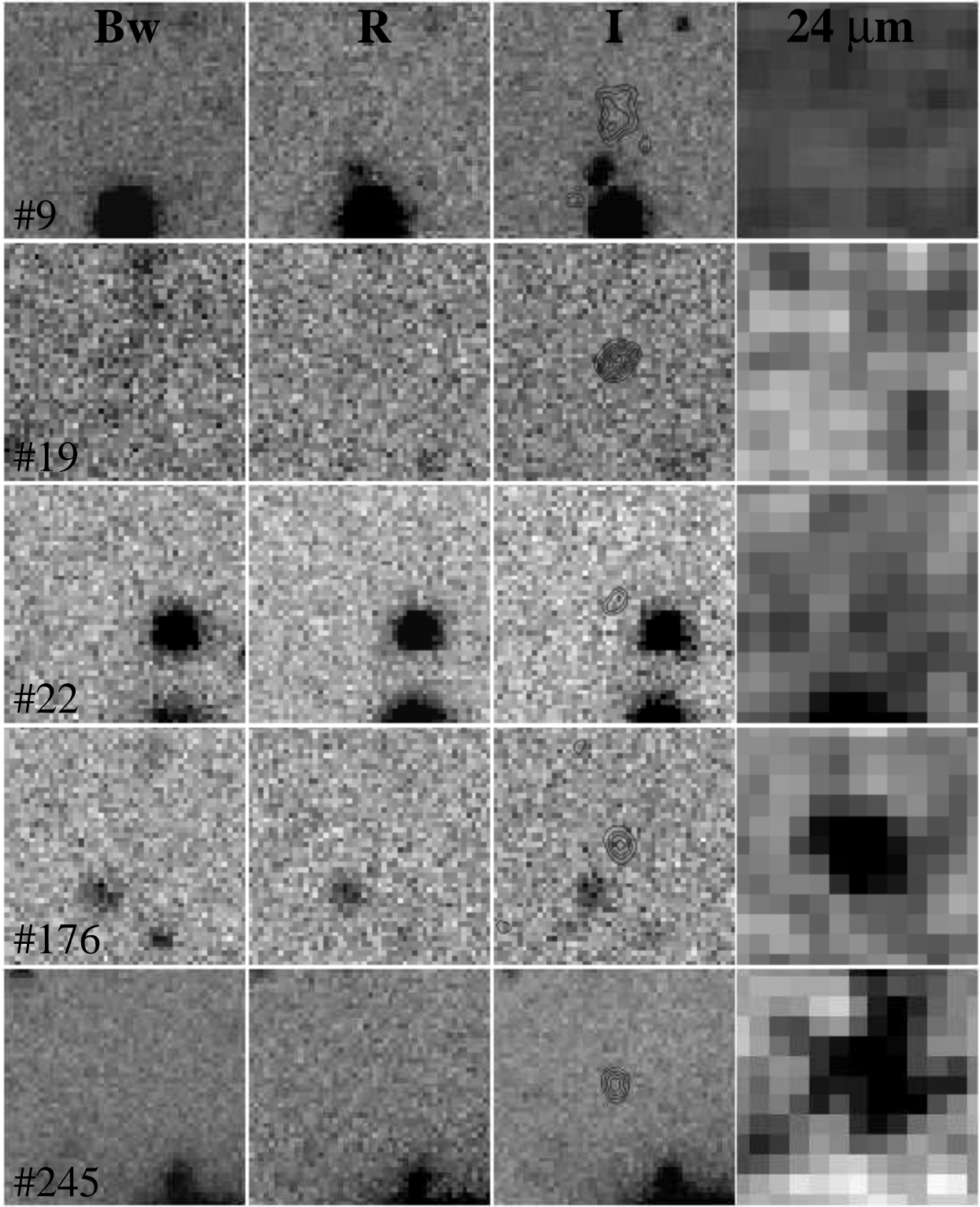}
\caption{ A sample of five optically invisible radio sources (OIRSs)
shown in optical $B_W$, $R$, $I$ and MIPS 24 $\mu$m bands. Radio contours
are shown superposed on the I image, with the lowest contour representing
the local 3 $\sigma$ level. Each field is $12\arcsec \times 12\arcsec$. }
\end{figure}

\clearpage
\begin{figure}
\figurenum{3}
\plotone{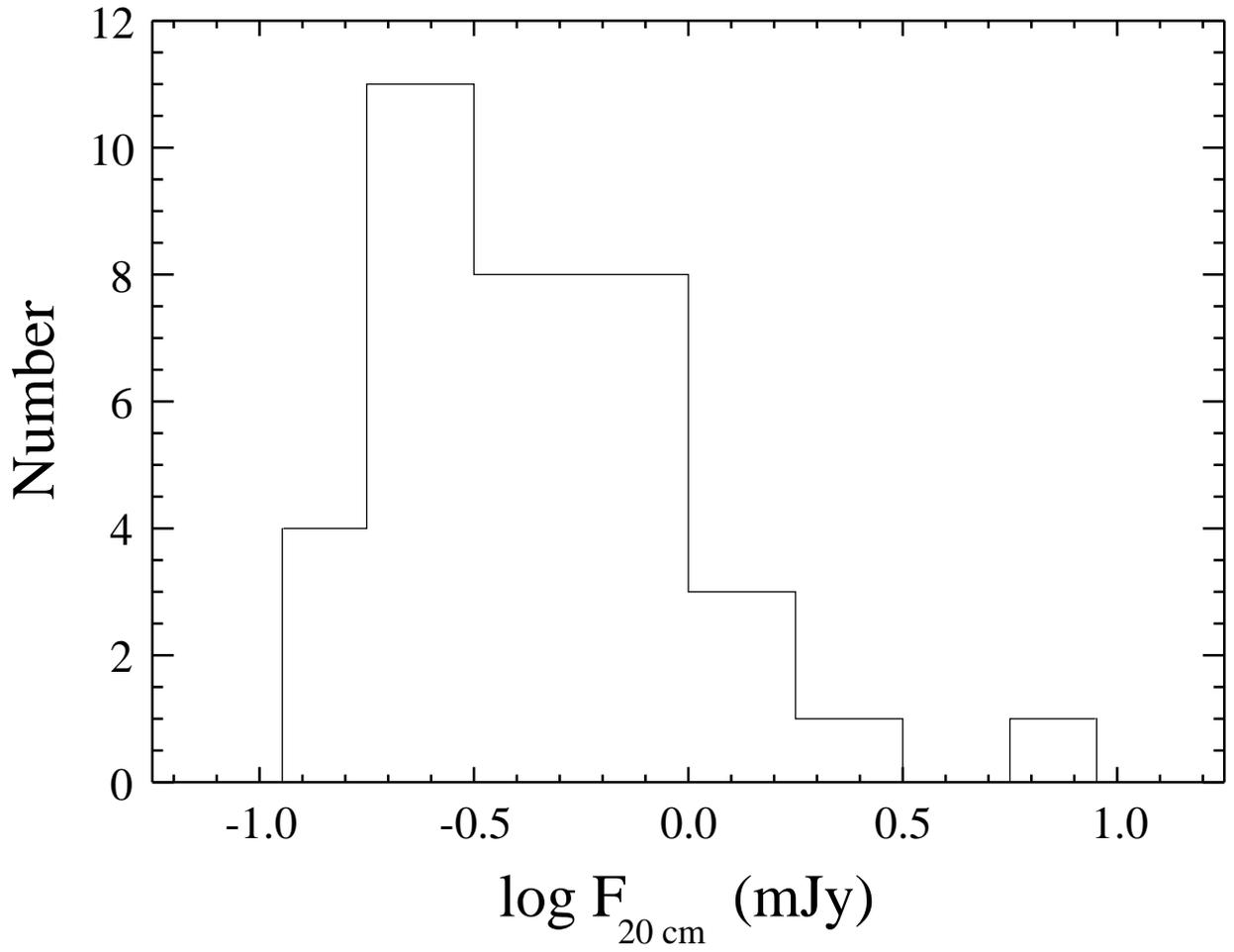}
\caption{Histogram of 20 cm flux densities for the OIRS sample.}
\end{figure}

\clearpage
\begin{figure}
\figurenum{4}
\plotone{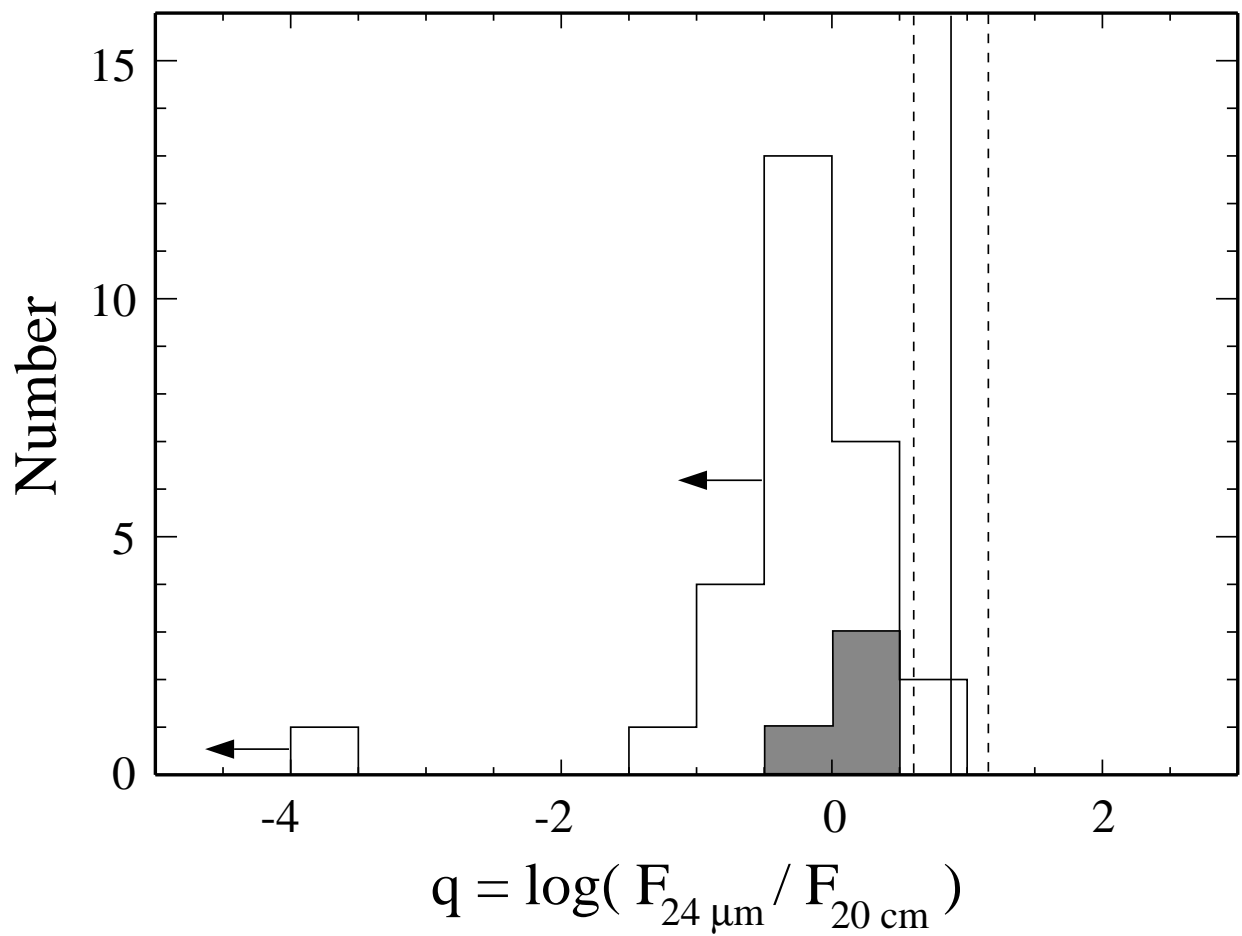}
\caption{Distribution of q for the OIRS sample. 
The shaded histogram represents the four OIRS detected
at 24 $\mu$m by MIPS. The
unshaded histogram represents upper-limits for q calculated using
F$_{\rm 24 \mu m}$ $<$ 0.3 mJy. The vertical solid and dashed lines represent 
respectively the mean value and range for q in z $\la$ 2 starburst dominated 
systems as determined by Appleton et al. (2004). }
\end{figure}

\clearpage
\begin{figure}
\figurenum{5}
\plotone{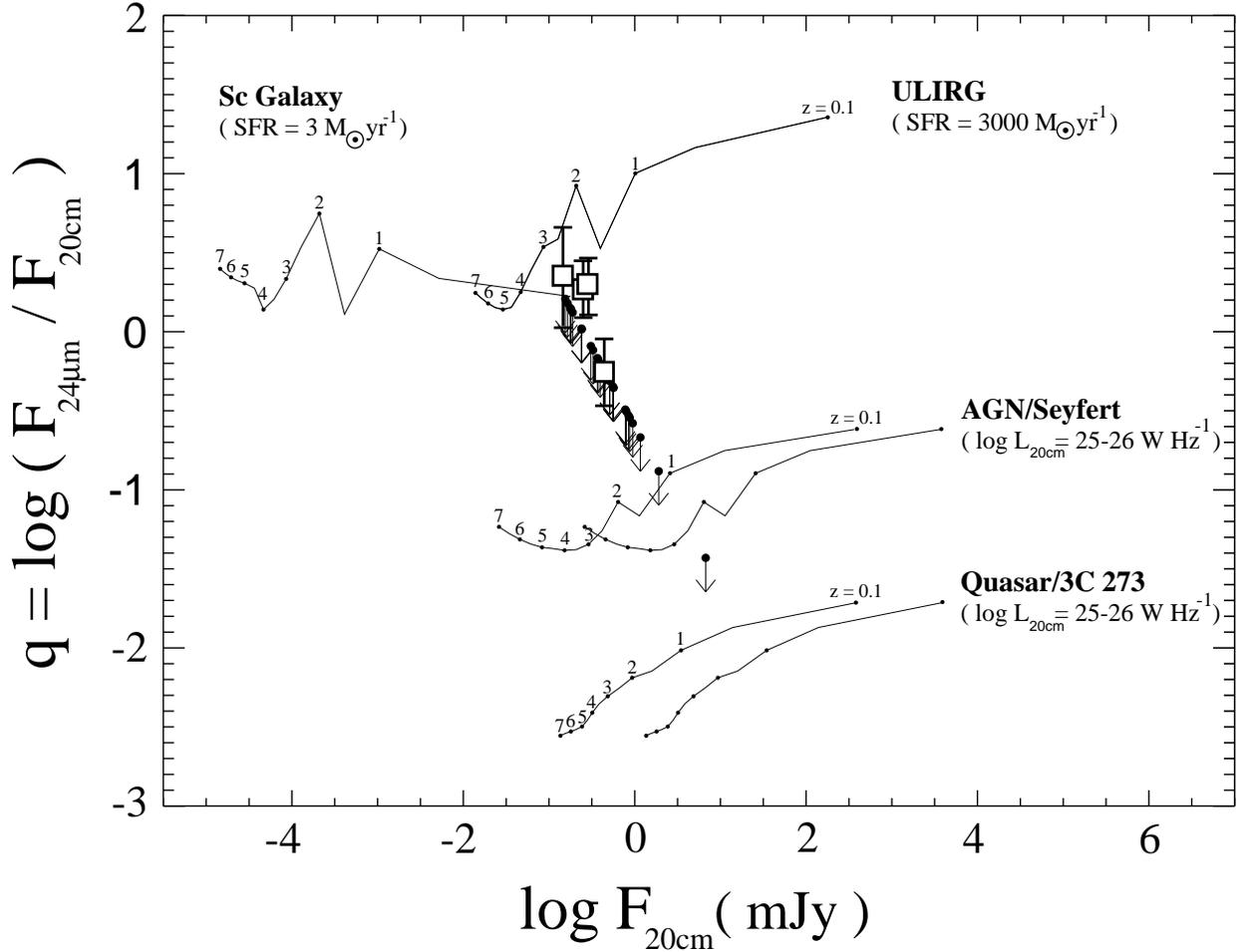}
\caption{ Infrared/Radio color-magnitude diagram.  The four sets of
tracks show the expected infrared to radio flux density ratio versus
radio flux density for various SEDs as a function of redshift. 
Top to bottom: a
SFR = 3 M$_{\odot}$ yr$^{-1}$ late spiral, a SFR = 3000 M$_{\odot}$
yr$^{-1}$ ULIRG, a Seyfert AGN with 20 cm luminosities of
10$^{25}$ and 10$^{26}$ W Hz$^{-1}$, and a quasar with 3C~273's SED and
20 cm luminosities identical to the Seyfert galaxy's.  Redshift is
indicated along the tracks.  The large unfilled squares represent the four
OIRSs detected at 24 $\mu$m. The small filled circles with
arrows are upper-limits for the remaining 27 OIRS not detected by MIPS.  }
\end{figure}

\clearpage
\begin{figure}
\figurenum{6}
\plotone{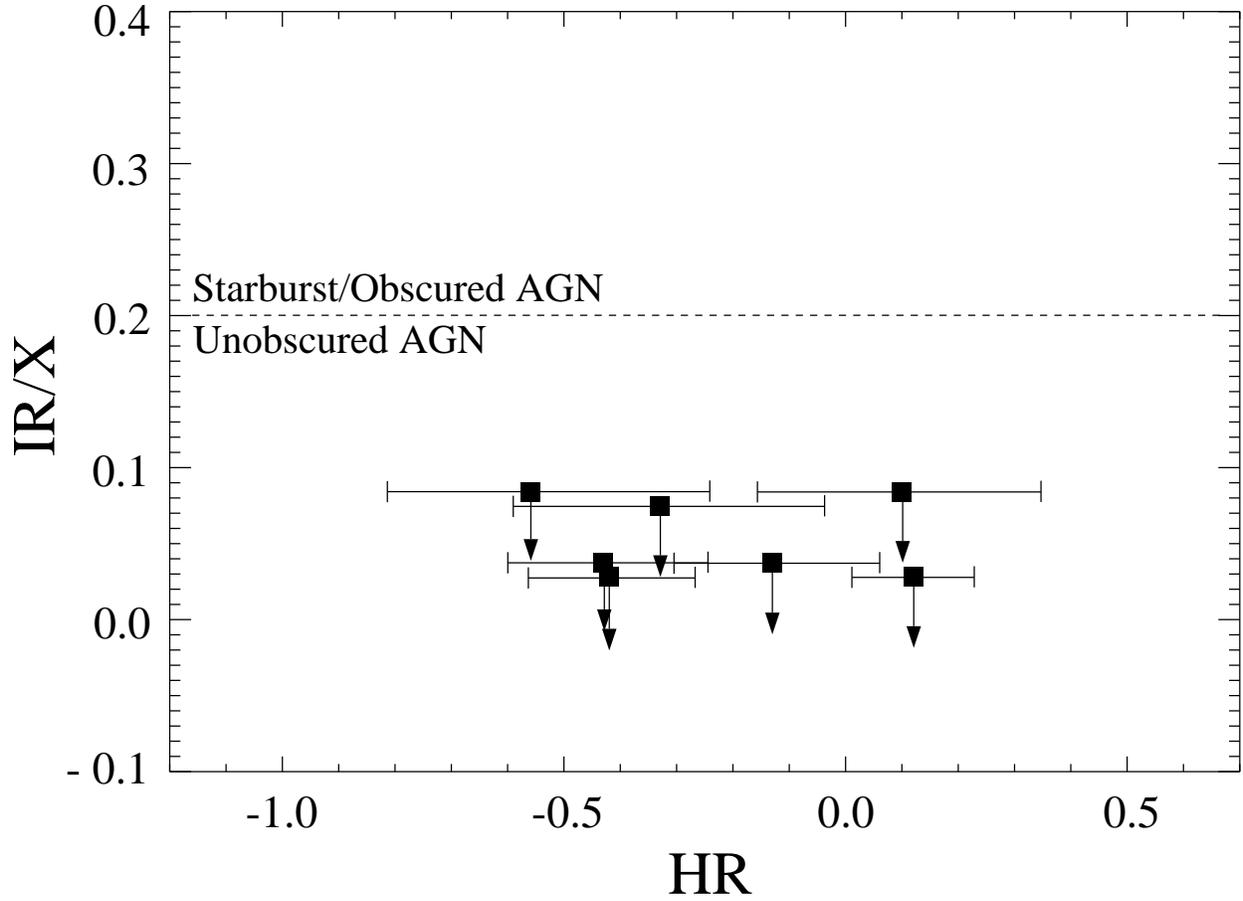}
\caption{The infrared to X-ray ratio (IR/X) versus
the X-ray Hardness ratio (HR) for the seven optically invisible
X-ray sources detected in the Chandra ``soft'' 0.5-2 keV band. 
The dashed line marks IR/X = 0.2, which separates starburst and 
obscured AGN powered emission (IR/X $>$ 0.2) from systems dominated 
by unobscured AGN (IR/X $<$ 0.2).}
\end{figure}

\clearpage
\begin{figure}
\figurenum{7}
\plotone{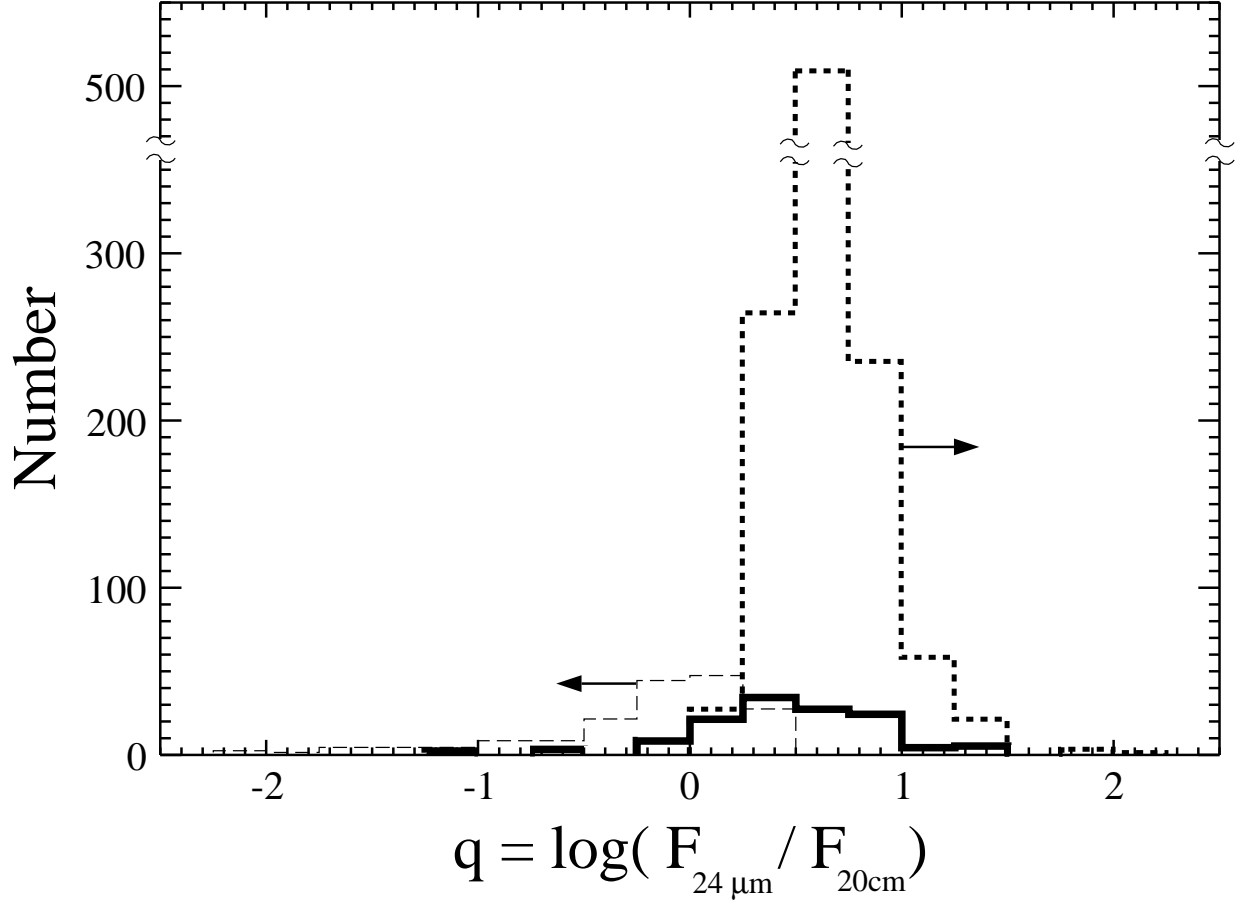}
\caption{Distribution of q for the full radio and infrared sample.
The thick solid-line histogram depicts q for the 125 sources 
detected at both 24 $\mu$m and 20 cm. The thick-dashed line histogram shows
q lower-limits for the 1086 sources detected only in the infrared, 
where the F$_{\rm 20cm}$ upper-limit was taken to be a point source
with 5$\times$ the rms at the sources position in the final radio image.
All sources in the histogram lie within or to the right of the bins shown. 
The thin-dashed line histogram shows the 194 radio source's not detected 
by MIPS, assuming a 24 $\mu$m flux upper-limit of 0.3 mJy.  Similarly, the 
sources in this histogram will lie within or to the left of the bins indicated.}
\end{figure}

\clearpage
\begin{figure}
\figurenum{8}
\plotone{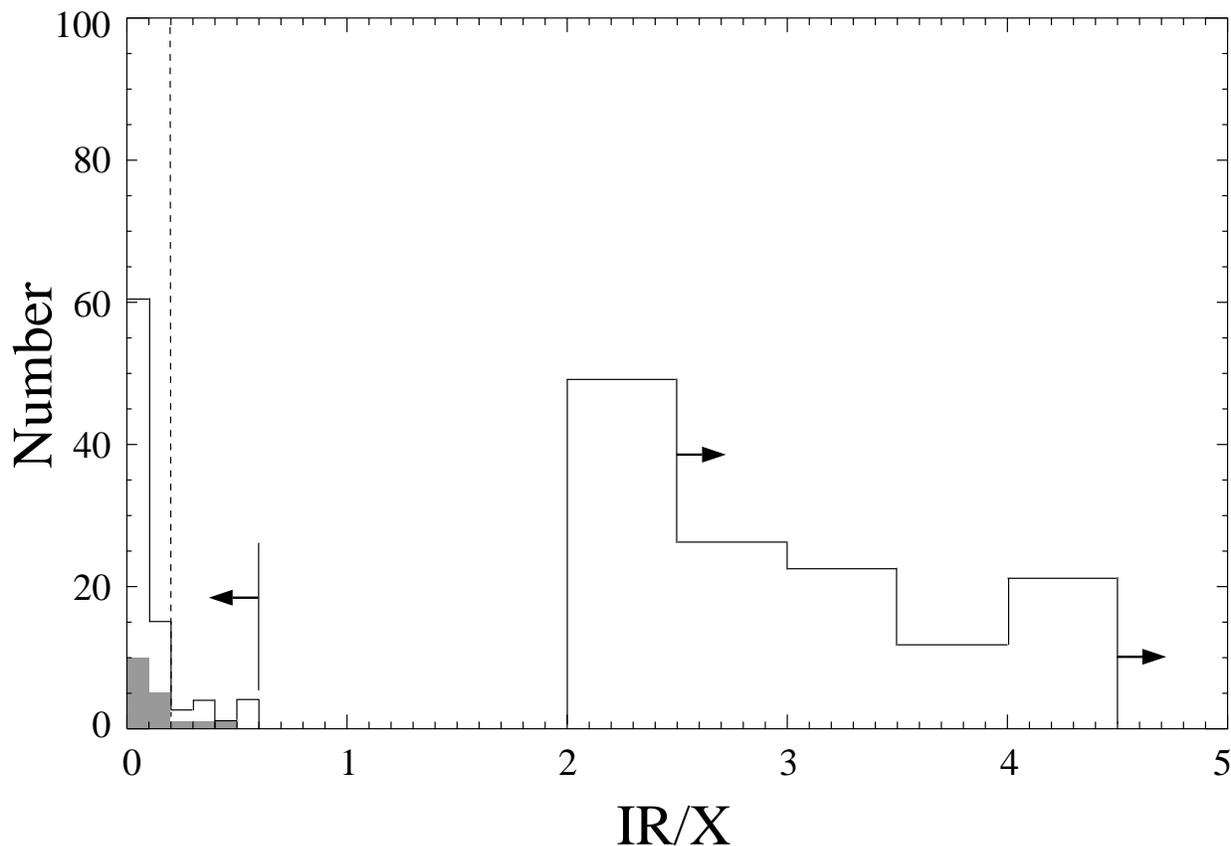}
\caption{The distribution of IR/X for the full
sample of infrared and X-ray sources in the overlapping survey area.
The shaded histogram shows values of IR/X for the 18 objects detected
in both MIPS 24 $\mu$m and Chandra 0.5-2 keV band emission.  The 
unshaded histograms show upper and lower limits to IR/X for the
Chandra-only (left histogram) and MIPS-only (right histogram) detected 
samples, respectively. Sources with IR/X $>$ 0.2 are consistent with
starbursts or heavily obscured AGN, which includes all MIPS-only sources. 
Sources with IR/X $<$ 0.2 have their luminosity dominated by an relatively 
unobscured AGN. This encompasses most Chandra-only sources.}
\end{figure}

\end{document}